\documentclass[letterpaper,english,aps,prd,floatfix,11pt,onecolumn]{revtex4}
\usepackage{geometry}
\usepackage{amsmath,amssymb,lmodern}
\usepackage{graphicx}
\usepackage{float}
\usepackage{subfig}
\usepackage{hyperref}
\usepackage{babel}

\def	\be	{\begin{equation}}
\def	\ee	{\end{equation}}
\def	\bqt	{\begin{quote}}
\def	\eqt	{\end{quote}}

\def	\del	{\nabla}

\begin{document}
\raggedbottom
\pagestyle{plain}
\pagenumbering{arabic}

\title{Einstein's Equations from the Stretched Future Light Cone}
\author{Maulik Parikh}
\author{Andrew Svesko}
\affiliation{Department of Physics and Beyond: Center for Fundamental Concepts in Science\\
Arizona State University, Tempe, Arizona 85287, USA}

\begin{abstract}
\noindent
We define the stretched future light cone, a timelike hypersurface composed of the worldlines of radially accelerating observers with constant and uniform proper acceleration. By attributing temperature and entropy to this hypersurface, we derive Einstein's equations from the Clausius theorem. Moreover, we show that the gravitational equations of motion for a broad class of diffeomorphism-invariant theories of gravity can be obtained from thermodynamics on the stretched future light cone, provided the Bekenstein-Hawking entropy is replaced by the Wald entropy.
\end{abstract}

\maketitle
\thispagestyle{empty}

\newpage

\section{Introduction}
\noindent
In the laws of black hole mechanics \cite{bardeen}, the area and surface gravity of a black hole event horizon are associated with entropy and temperature. These laws point to a relation between classical geometry and thermodynamics, using global equations applicable to stationary spacetimes that contain black holes. However, the fact that de Sitter and Rindler horizons --- which are observer-dependent and therefore could be anywhere --- also have thermodynamic properties suggests that holographic entropy and temperature are actually more generally applicable concepts in spacetime. Taking this idea significantly further, Jacobson \cite{Jacobson:1995ab} attributed thermodynamic properties even to local Rindler horizons, which are essentially just planar patches of certain null congruences passing through arbitrary points in spacetime, and are not event horizons in any global sense. The locality of local Rindler ``horizons" has the effect that local equations follow from thermodynamic equations. Specifically, Einstein's equations follow from the Clausius theorem, $Q = T \Delta S$; more recently \cite{thermoNEC,tworoads}, the null energy condition has been obtained from the second law of thermodynamics.

Here we present a new formulation: we attribute thermodynamic properties to the future light cone of any point, $p$, in an arbitrary spacetime. A future light cone can be regarded as a kind of spherical Rindler horizon because the worldlines of observers with constant outward radial acceleration asymptote to it. In fact, it will be more convenient to consider the stretched future light cone, a timelike codimension-one hypersurface. Indeed, we will define our stretched future light cone as a timelike congruence of worldlines with approximately constant and uniform radial acceleration. By constant, we mean that the proper acceleration of any single worldline does not change along the worldline; by uniform, we mean that all worldlines share the same proper acceleration. 

Given the relation between temperature and acceleration, it then seems natural to attribute a constant and uniform temperature to this surface. In fact, entropy is also a somewhat better-motivated property of our surface than of local Rindler horizons. This is because a future light cone separates its interior from the exterior spacetime; the interior is causally disconnected from the exterior, in the same sense that the interior of a black hole is. It seems therefore plausible that we might associate entropy to spacelike sections of the light cone, for example as the entanglement entropy between the interior and exterior regions. By contrast, a finite strip of Rindler horizon (unlike an infinite global Rindler horizon) does not separate space into two disconnected regions, and it is not obvious that it should possess an entropy. Another appealing feature of our formulation is that the interior of a future light cone resembles that of black holes or de Sitter space in that it admits compact spatial sections. 

These geometric aspects motivate the premise of this paper, which is that holographic thermodynamic properties can be associated locally with the stretched future light cone emanating from an arbitrary point $p$ in an arbitrary spacetime. We will then show that the Clausius theorem, properly understood, yields Einstein's equation at $p$,
\be
Q = T \Delta S \Rightarrow R_{ab} - \frac{1}{2} R g_{ab} + \Lambda g_{ab} = 8 \pi G T_{ab} \; ,
\ee
much as the association of thermodynamics with local Rindler horizons leads to Einstein's equation emerging as an equation of state \cite{Jacobson:1995ab}.

Besides its conceptual appeal, the stretched future light cone formulation of local holographic thermodynamics also offers a significant new result: it permits the extension of Jacobson's result to a wide class of theories of gravity. It has been a longstanding challenge to obtain the gravitational equations of motion for general, higher-curvature theories of gravity from  thermodynamics. Broadly, we can divide earlier attempts into two categories: (i) those that aim to derive the equations of motion for $f(R)$ theories of gravity via a nonequilibrium modification of the Clausius theorem to account for internal entropy production terms \cite{jacobsonf(R),Cai,ElizaldeSilva}, and (ii) those that aim to derive the gravitational equations for general theories of gravity \cite{Parikh:2009qs,Brustein:2009hy,Padmanabhan:2009ry,Guedens:2011dy,Dey:2016zka}. The approaches that fall into category (i) have been critically reviewed in \cite{Guedens:2011dy}, which points out that this nonequilibrium approach can never lead to theories beyond $f(R)$ gravity. The attempts that fall into category (ii) mainly use a ``Noetheresque" approach, in which the local entropy is expressed as an integral of a Noether current \cite{Parikh:2009qs,Brustein:2009hy,Guedens:2011dy,Dey:2016zka} over spacelike sections of a local Rindler plane. Unfortunately, all the early papers using the Noetheresque approach contained technical errors, as reviewed in \cite{Guedens:2011dy}. Although the authors of \cite{Guedens:2011dy} fixed the technical problems, the derivation nonetheless appears quite unphysical, with the entropy not always proportional to the area even for Einstein gravity. The present work applies the Noetheresque approach of Parikh and Sarkar \cite{Parikh:2009qs} to the setting of a stretched future light cone, rather than to local Rindler planes. As we shall see, the geometry of the new setup allows the technical problems in earlier derivations to be overcome while still preserving an entropy proportional to the area for Einstein gravity. We will describe the earlier literature of the Noetheresque approach, as well as its technical challenges, in more detail in Sec. IV.

In this work, we consider those gravitational theories whose Lagrangian consists of a polynomial in the Riemann tensor (with no derivatives of the Riemann tensor, for simplicity). For all such theories, after replacing the Bekenstein-Hawking entropy with the Wald entropy, we find that Clausius' theorem again implies the field equations of classical gravity:
\be
Q = T \Delta S \Rightarrow P_{a}^{\;\;cde}R_{bcde}-2\nabla^{c}\nabla^{d}P_{acdb}-\frac{1}{2}Lg_{ab}=8\pi G T_{ab} \; ,
\ee
where the equation on the right is, as we shall describe, the generalization of Einstein's equations for these higher-curvature gravitational theories, up to an undetermined cosmological constant term.

In summary, the main goals of this paper are, first, to formulate a definition of the stretched future light cone and, second, to derive the (generalized) Einstein equations from the premise that local holographic thermodynamic properties can be attributed to stretched future light cones.

\section{Construction}
\noindent
Our first task is to carefully define what we mean by a stretched future light cone. We also need to be precise in defining its thermodynamic properties. We begin by adapting the notion of approximate Killing vectors for the construction of spherical Rindler horizons.

\subsection{Approximate Killing Vectors}

In the vicinity of any point, $p$, spacetime is locally flat. Components of the metric tensor can therefore be expanded in Riemann normal coordinates:
\be
g_{ab} (x) = \eta_{ab} - \frac{1}{3} R_{acbd} (p) x^c x^d + ... \; ,\label{RNC}
\ee
where the Riemann tensor is evaluated at the point $p$, which lies at the origin of the Riemann normal coordinate system. Here the $x^a$ are Cartesian coordinates and $\eta_{ab}$ is the Cartesian Minkowski metric; in Riemann normal coordinates, the Christoffel symbols vanish at $p$ and the metric expansion has no piece that is linear in $x$. 

The local flatness of spacetime means that there exist $D$-choose-two independent vectors $\xi^a$ in the tangent plane, $T_p$, which are the Killing vectors of $D$-dimensional Minkowski space, and correspond to local translations and local Lorentz symmetries. When spacetime is not exactly Minkowski space, these vectors are not exactly Killing vectors; call them approximate Killing vectors. More precisely, in a generic spacetime, the presence of quadratic terms of ${\cal O}(x^2)$ in the Riemann normal coordinate expansion, Eq. (\ref{RNC}), indicates that Killing's equation for these vectors will fail at some order in $x$. The order depends on the nature of the approximate Killing vector: for translations the components of the Killing vector are constants, whereas for Lorentz transformations, $x_{\mu} \partial_{\nu}^a - x_{\nu} \partial_{\mu}^a$, the components themselves are of ${\cal O}(x)$. Thus for the generators of local Lorentz transformations, Killing's equation fails in a generic spacetime at ${\cal O}(x^2)$. Note also that Killing's identity,
\be
\nabla_{a}\nabla_{b}\xi_{c}=R^{d}_{\;abc}\xi_{d} \; ,
\ee
which is a consequence of Killing's equation, fails for these vectors at ${\cal O}(x)$. That is, we have
\be 
\nabla_{a}\xi_{b}+\nabla_{b}\xi_{a}\approx\mathcal{O}(x^{2}) \; , \label{boostKilling}
\ee
and
\be 
\nabla_{a}\nabla_{b}\xi_{c}-R^{d}_{\;abc}\xi_{d}\approx\mathcal{O}(x) \; , \label{boostfailure}
\ee
for approximate Killing vectors generating local Lorentz transformations.

Now, the integral curves (flow lines) of Cartesian boosts trace the worldlines of Rindler observers -- observers with constant acceleration in some Cartesian direction. Here, however, we are interested in considering a congruence of observers that sweep out a stretched future light cone. Regarding the future light cone as a spherical Rindler horizon, we are motivated to define the stretched future light cone as a congruence of worldlines generated by spherical boosts. Hence we define $\xi_a$ as follows:
\be
\xi_a \equiv -r \delta_{ta} + t \delta_{ra} = -\sqrt{x_i x^i} \delta_{ta} + \frac{t}{\sqrt{x_ix^i}} x^j \delta_{ja} \; , \label{xi}
\ee
where $r$ is the radial coordinate while $x^i$ are spatial Cartesian coordinates, in some split of spacetime into space and time. (In the Appendix, we will refine this somewhat by allowing $\xi_a$ to have small sub-leading modifications that are quadratic and higher in Riemann normal coordinates, with constant coefficients that depend on the Riemann tensor and its derivatives at $p$, these subleading terms, which vanish in Minkowski space, will play a useful role in our derivation of the field equations.)

Note that $\xi_a$ is not a Killing vector. This is because $\xi_a$ generates radial boosts but radial boosts are not isometries even of Minkowski space. More precisely, the symmetric covariant derivatives $\nabla_a \xi_b + \nabla_b \xi_a$ are
\be
\begin{split}
&\nabla_{t}\xi_{t}=0 + {\cal O}(x^2)\,,\quad \nabla_{t}\xi_{i}+\nabla_{i}\xi_{t}=0 + {\cal O}(x^2)\,,\\
&\nabla_{i}\xi_{j}+\nabla_{j}\xi_{i}=\frac{2t}{r}\left(\delta_{ij}-\frac{x_{i}x_{j}}{r^{2}}\right) + {\cal O}(x^2) \; .	\label{Killingfailure}
\end{split}
\ee	
Notice that the $t-t$ and $t-i$ components satisfy Killing's equation at ${\cal O}(1)$ whereas the $i-j$ components fail to obey Killing's equation even at that leading order. (In spherical coordinates, the $i-j$ terms correspond to angle-angle components of the symmetric covariant derivatives.) The ${\cal O}(x^2)$ corrections generically appear from Christoffel symbols multiplying the linear pieces of $\xi_a$, as in (\ref{boostKilling}). 

\subsection{Definition of the stretched future light cone}

We are now ready to define the stretched future light cone. To gain some intuition, let us first define the stretched future light cone in Minkowski space. As in (\ref{xi}), define
\be
\xi_a^{\rm Mink} \equiv -r \delta_{ta} + t \delta_{ra} \; .
\ee
The flow lines of $\xi_{\rm Mink}a$ trace out hyperbolas. Define a codimension-one timelike hyperboloid by the set of curves that obey
\be
r_{\rm Mink}^2 - t^2 = \alpha^2 \; , \label{hyperboloid}
\ee
where $t \geq 0$ and $\alpha$ is some given scale with dimensions of length. In Minkowski space, this hyperboloid is a stretched future light cone because, as $t \rightarrow +\infty$, it asymptotes to the future light cone emanating from the point $p$ at the origin. In $D$-dimensional spacetime, the constant-$t$ sections of the hyperboloid are $D-2$-dimensional spheres with area
\be
A_{\rm Mink}(t) = \Omega_{D-2} (\alpha^2 + t^2)^{\frac{D-2}{2}} \; . \label{areamink}
\ee
On this hyperboloid, we have
\be
\xi_{\rm Mink}^2 = - \alpha^2 \; . \label{ximinkmag}
\ee
We can regard $\xi_a$ as the unnormalized tangent vector to the worldlines of our Rindler observers. These have normalized velocity vector
\be
u_a^{\rm Mink} \equiv \frac{\xi^{\rm Mink}_a}{\alpha} \; ,
\ee
where $u^2 = -1$. The proper acceleration of such observers, $a_{\rm Mink}^c \equiv u_{\rm Mink}^b \nabla_b u_{\rm Mink}^c$, has magnitude
\be
a_{\rm Mink} = \frac{1}{\alpha} \; . \label{xiacc}
\ee
The hyperboloid therefore is a congruence of worldlines of a set of constant radially accelerating observers all with the same uniform acceleration $1/\alpha$. 

Now let us think about how to define our stretched future light cone when $p$ lies in a general curved spacetime. In Minkowski space, the locus of points defined by (\ref{hyperboloid}), (\ref{ximinkmag}), and (\ref{xiacc}) are all the same. However, in curved spacetime, these three expressions are no longer equivalent. A straightforward calculation shows that
\be
\xi^2 = - \alpha^2 + {\cal O}(x^4) \label{ximag}
\ee
and
\be
a = \frac{1}{\alpha} \left (1 + {\cal O}(x^4) \right ) \label{generalacc}
\ee
How then should we choose our stretched future light cone? (A previous proposal \cite{Kothawala:2014oba} considered equigeodesic surfaces, the locus of points a fixed finite geodesic distance from $p$. Although such surfaces agree with the hyperboloid in Minkowski space, this is not how we will define our stretched future light cone in a general curved spacetime.) Our choice is motivated by the stretched horizon of the black hole membrane paradigm, which is a congruence of the worldlines of fiducial observers.  Call our stretched future light cone $\Sigma$. Since we are interested in thermodynamics, we would like $\Sigma$ to be a surface of constant and uniform temperature. Then, since temperature is related to acceleration, we would like our surface to be composed of a congruence of timelike worldlines of constant proper acceleration; a similar construction was proposed by Piazza \cite{Piazza:2010hz}. That suggests using $a = 1/\alpha$ as our definition of $\Sigma$. However, there is a slight problem: as a result of spacetime curvature, none of the flow lines of (\ref{xi}) typically correspond to worldlines with constant acceleration.

We therefore define $\Sigma$ as follows. First, pick a small length scale, $\alpha$. By small, we mean that the metric should be roughly flat to a coordinate distance $\alpha$ from the origin of Riemann normal coordinates or that $\alpha$ is much smaller than the smallest curvature scale at $p$. Next, imagine that the radial boost vector field $\xi_a$, as defined by (\ref{xi}), consists of the (unnormalized) tangent vectors to the worldlines of a set of observers. Among this set, select the subset of observers who, at time $t = 0$, have instantaneous proper acceleration $1/\alpha$. (If spacetime were flat, this subset of observers would describe a codimension-two sphere of radius $\alpha$ at $t = 0$, as given by (\ref{hyperboloid}). However, since spacetime is not exactly flat, the subset forms a codimension-two surface $\omega(0)$, which is a small deformation of the $r = \alpha$ surface; that deformation will play no further role.) Now, as already mentioned, if we were to follow the worldlines of these observers, they would generically not have the same proper acceleration $1/\alpha$ at some later time. To avoid this problem, choose a timescale $\epsilon$. If $\epsilon$ is very short,
\be
\epsilon \ll \alpha \; ,
\ee
then we can regard the proper acceleration of our initially accelerating observers to be approximately constant over that timescale. We therefore restrict our calculations to the range
\be
0 \leq t \leq \epsilon \; .
\ee
Over this interval, we can regard our stretched future light cone $\Sigma$ to be the world tube of a congruence of observers with the same nearly constant approximately outward radial acceleration $1/\alpha$ (Fig. 1).

\begin{figure}[H]
\centering
 \includegraphics[width=.75\textwidth]{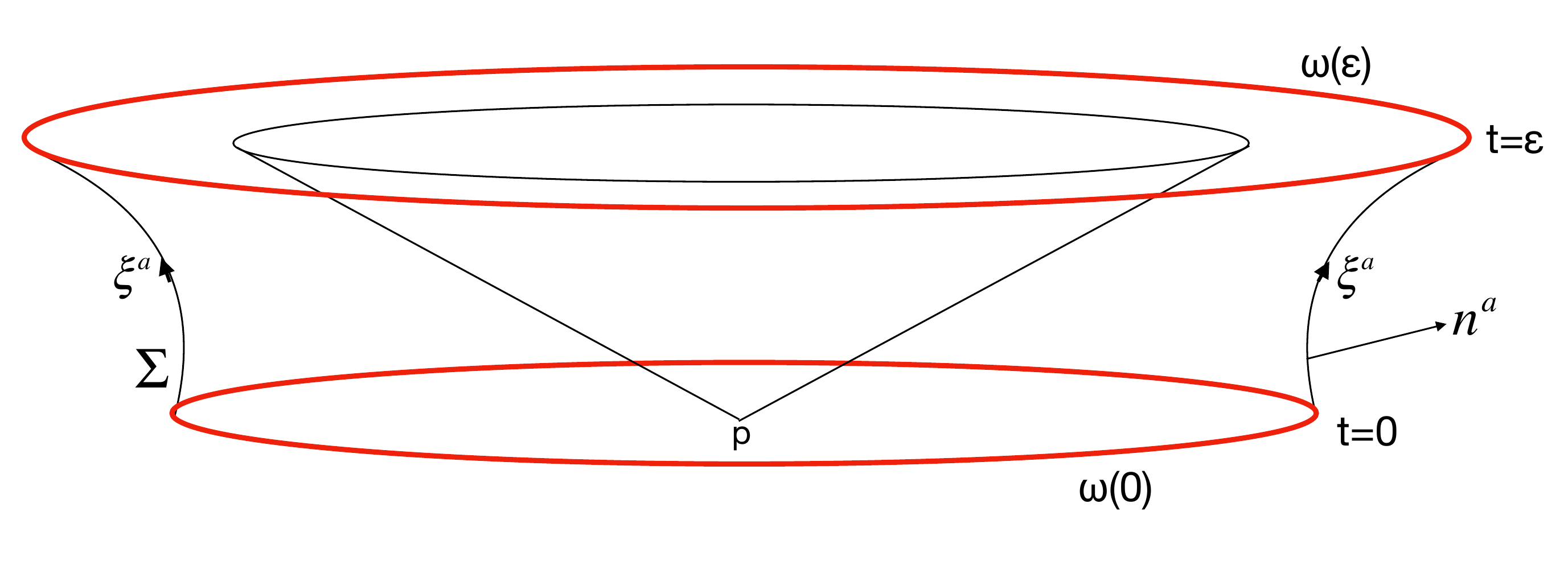}
 \caption{A congruence of radially accelerating worldlines $\xi^{a}$ with the same uniform proper acceleration $1/\alpha$ generates the stretched future light cone of $p$, and describes a timelike hypersurface, $\Sigma$, with unit outward-pointing normal $n^a$. The boundary of $\Sigma$ consists of the two codimension-two surfaces $\omega(0)$ and $\omega(\epsilon)$ given by the constant-time slices of $\Sigma$ at $t=0$ and $t=\epsilon$, respectively.
}
 \end{figure}
 
The overall effect of spacetime curvature is to make $\Sigma$ a small deformation of the hyperboloid $r^2 = \alpha^2 + t^2$, and to restrict the time interval to the range $0 \leq t \leq \epsilon \ll \alpha$. From (\ref{ximag}), the normalized velocity vectors are
\be
u_a \equiv \frac{\xi_a}{\sqrt{-\xi^a \xi_a}} \approx \frac{\xi_a}{\alpha} \; , \label{velocity}
\ee
while the normal to $\Sigma$ is a small correction to the normal to the hyperboloid:
\be
n_a \approx -\frac{t}{\alpha} \delta_{ta} + \frac{r}{\alpha} \delta_{ra} + \dots \; .
\ee
The proper acceleration of our observers is
\be
a^b = u^a \nabla_a u^b = \frac{1}{\alpha} n^b \label{properacc}
\ee
and has magnitude $1/\alpha$ on $\Sigma$.

The reason for choosing $\Sigma$ to be a hypersurface composed of constant acceleration worldlines is that, by the relation between temperature and acceleration, $\Sigma$ then becomes an isothermal surface. A rigorous identification of temperature with acceleration follows from the choice of a Poincar\'e-invariant vacuum state. The existence of an approximately Poincar\'e-invariant vacuum state is a consequence of the strong principle of equivalence. If we assume that free-falling observers should see the same physics locally as inertial observers in Minkowski space, then we are naturally led to assume that the quantum state responsible for local physics should be approximately the Poincar\'e-invariant state of Minkowski space; any other coherent state would have a stress tensor whose vacuum expectation value would be singular somewhere. The same prescription is used to select the Unruh state in the black hole case, ensuring that an observer falling along a geodesic sees no Hawking radiation. The validity of using the Poincar\'e-invariant state locally even has experimental support in that high-energy physics at accelerators is perfectly captured by quantum field theory in Minkowski space, even though on larger scales our spacetime is not well described by Minkowski space. 

Given the Poincar\'e-invariant vacuum state, we automatically find that the expectation value of the Rindler number operator is thermal; the state is thermal with respect to generators of Lorentz boosts. Operationally, this means that eternally accelerating Rindler observers equipped with Unruh detectors will detect particles with a thermal spectrum. Transient acceleration in Minkowski space was studied by Barbado and Visser \cite{Barbado:2012fy} who found that a thermal spectrum is still detected provided the duration of acceleration is sufficiently long compared with the inverse acceleration. This condition is easy to arrange in our construction. We need to extend the worldlines of the accelerating observers over a longer time, $\tau$, much greater than the inverse acceleration, $\alpha$ (but still short enough that curvature effects are negligible). Since there is no limit to how small $\alpha$ can be, we can always do this. Our surface $\Sigma$ is then a brief segment, $0 < t < \epsilon \ll \alpha \ll \tau$ of a more extended surface traced by a congruence of such observers. Observers who continue to accelerate on the surface beyond $\Sigma$ with the same constant acceleration $1/\alpha$ will detect a roughly thermal spectrum whose temperature matches their acceleration. In general, the worldlines of the observers will not be integral curves of our approximate Killing vector $\xi_a$ before $t = 0$ or after $t = \epsilon$. We therefore restrict our calculation to $\Sigma$ because we need a congruence generated by the flow lines of $\xi_a$. 

The existence of an approximately Poincar\'e-invariant state therefore ensures that $\Sigma$ is an isothermal surface with Davies-Unruh temperature
\be
T \equiv \frac{\hbar a}{2 \pi} = \frac{\hbar}{2 \pi \alpha} \; . \label{temp}
\ee
In particular, this means that in any integration over $\Sigma$, we can move the temperature outside the integral.

\subsection{Definition of $S$}

Having defined our stretched future light cone, $\Sigma$, and having associated a uniform temperature with it, we next need to define the entropy. The underlying premise of the ``thermodynamics of spacetime" is that gravitational entropy can be attributed not just to global event horizons, but also to local Rindler horizons. In the same vein, we attribute a local entropy to spacelike sections of the future light cone \cite{DeLorenzo:2017tgx}. We also attribute entropy to sections of our timelike stretched horizon, $\Sigma$. This is consistent with the black hole membrane paradigm in which the timelike stretched horizon can also be thought of as having thermodynamic properties \cite{membrane}. 

The form of the entropy depends on the gravitational theory under consideration. For Einstein gravity, the entropy is the Bekenstein-Hawking entropy, one quarter of the area measured in Planck units:
\be
S = \frac{A}{4 G \hbar} \; . \label{Bek}
\ee
We will first rewrite this in a useful form using the vectors $n_{a}$ and $\xi_{a}$ on $\Sigma$. Let $\omega(t)$ be the codimension-two section of $\Sigma$ at time $t$.
Its area is
\be 
A(t) \equiv \int_{\omega(t)} dA = \alpha \int_{\omega(t)} dA \, n_b \frac{1}{\alpha} n^b =  \alpha \int dA \, n_b u^a \nabla_a u^b = \int dA \, n_b u_a \nabla^a \xi^b \; .
\ee
Here we have used (\ref{velocity}) and (\ref{properacc}).
Next we make use of the fact that $\nabla_{a}\xi_{b}=-\nabla_{b}\xi_{a}$ for the projection of $\nabla_{a}\xi_{b}$ in the $n-\xi$ plane, as we see from the first line of (\ref{Killingfailure}). Then defining
\be
dS_{ab} \equiv \frac{1}{2}(n_{a}u_{b}-n_{b}u_{a})dA \; ,
\ee
we see that the Bekenstein-Hawking entropy at time $t$ can be expressed as 
\be
S(t) = - \frac{1}{4G \hbar} \int_{\omega(t)} dS_{ab} \del^a \xi^b = - \frac{1}{4G \hbar} \int_{\omega(t)} dS_{ab} \frac{1}{2} (g^{ac} g^{bd} - g^{ad} g^{bc} ) \del_c \xi_d \; . \label{areaxi}
\ee
Here we have written the entropy in the form $\int dS_{ab} M^{ab}$, where $M^{ab}$ is an antisymmetric tensor; this form will be helpful in deriving Einstein's equations and will generalize readily to other theories of gravity.

\section{Einstein's Equations}
\noindent
Now let us calculate the total change in the Bekenstein-Hawking entropy $\Delta S_{\rm tot} = S(\epsilon) - S(0)$, between $t = 0$ and $t = \epsilon$. To that end, note that the codimension-two surfaces $\omega(\epsilon)$ and $\omega(0)$ are the boundaries of the stretched future light cone, $\Sigma$ (Fig. 1). We can therefore make use of Stokes' theorem for an antisymmetric tensor field $M^{ab}$,
\be
\int_{\Sigma}d\Sigma_{a}\nabla_{b} M^{ab}=-\int_{\omega(\epsilon)}dS_{ab} M^{ab} + \int_{\omega(0)}dS_{ab} M^{ab} \; , \label{stokesthm}
\ee
where the overall minus sign arises because $\Sigma$ is a timelike surface. From (\ref{areaxi}), we find
\be
\Delta S_{\rm tot}=\frac{1}{4G \hbar}\int d\Sigma_{a}\frac{1}{2}(g^{ac}g^{bd}-g^{ad} g^{bc})(R^{e}_{\;bcd}(p) \xi_{e}+f_{bcd}) \label{DStot}
\ee
where we have approximated the Riemann tensor by its value at the point $p$, which we can do to leading order in $x$. To obtain (\ref{DStot}), we have written the Killing identity for our approximate Killing vector $\xi_a$ as
\be
\nabla_{b}\nabla_{c}\xi_{d}=R^{e}_{\;bcd}\xi_{e}+f_{bcd} \; . \label{fbcd}
\ee

The term $f_{bcd}$ accounts for the failure of Killing's identity to hold; for a true Killing vector, $f_{bcd}$ would be zero. As we see from (\ref{Killingfailure}), $\xi_a$ fails to obey Killing's equation in two ways. First, because of spacetime curvature, Killing's equation generically fails at quadratic order in Riemann normal coordinates. These quadratic terms contribute terms of order $x$ to $f_{bcd}$. But second, even if spacetime were exactly Minkowski space, our $\xi_a$ generates not planar boosts, but radial boosts; these are not true isometries, as indicated by the leading-order failure of Killing's equation to hold for the $i-j$ components. This contributes terms of order ${\cal O}(x^{-1})$ to $f_{bcd}$. (In addition to these, there will also be terms ${\cal O}(1)$ in $f_{bcd}$ coming from modifications to $\xi_a$, as detailed in Appendix A.) We cannot discard either of these pieces of $f_{bcd}$ because they are not higher order than the $R^{e}_{\;bcd}(p)\xi_{e}$ term we would like to keep, which is of order $x$. Fortunately, we do not need $f_{bcd}$ to vanish: as we shall see, we only need its integral to vanish. This distinction makes a tremendous difference. We note that because the constant-$t$ sections of $\Sigma$ are spheres (to leading approximation), any odd power of a spatial Cartesian coordinate $x^i$ integrates to zero over $\Sigma$. As shown in Appendix A this results in the vast majority of terms of order $x$ (and ${\cal O}(1)$) in $f_{bcd}$ integrating to zero. The handful of surviving terms can be canceled by including quadratic and cubic terms in the expansion of $\xi_a$. The same is not true for the term of order $1/x$ in $f_{bcd}$, which neither vanishes upon integration, nor can be canceled by redefinitions. To leading order, we can evaluate it in $D$-dimensional Minkowski space, where we find
\be
\frac{1}{4G \hbar}\int d\Sigma_{a}\frac{1}{2}(\eta^{ac}\eta^{bd}-\eta^{ad} \eta^{bc}) f^{{\cal O}(x^{-1})}_{bcd} =  \frac{\Omega_{D-2}}{4 G \hbar} \alpha^{D-4} \epsilon^2 \; . \label{nPf-1Einstein}
\ee
Remarkably, this term actually has a physical interpretation. 

Recall that we would like to equate our entropy change to the heat flux. However, as we have defined it, $\Delta S_{\rm tot}$ is the total change in the area of our stretched future light cone. Not all of this change in area can be attributed to the influx of heat. This is because $\Sigma$ is generated by a congruence of outwardly accelerating worldlines whose area would increase even in the absence of heat. Indeed, even in  Minkowski space with no heat flux whatsoever, the area of the hyperboloid of outwardly accelerating observers increases in time, Eq. (\ref{areamink}). Therefore, before identifying the change in entropy with $T^{-1} Q$, we should first subtract this background expansion of the hyperboloid, $\Delta S_{\rm hyp}$, from $\Delta S_{\rm tot}$:
\be
\Delta S_{\rm rev} \equiv \Delta S_{\rm tot} - \Delta S_{\rm hyp}	\label{revS}
\ee
We call the difference $\Delta S_{\rm rev}$, the reversible change in entropy, in analogue with ordinary thermodynamics for which we have $Q = T \Delta S_{\rm rev}$ (the general formula in the presence of irreversible processes is $\Delta S \geq Q/T$, with saturation only for the reversible component of $\Delta S$).

Now the change in the Bekenstein-Hawking entropy from the natural expansion of the stretched future light cone can be read off from (\ref{areamink}). It is
\be
\Delta S_{\rm hyp} = \frac{\Omega_{D-2}}{4 G \hbar} \left (r^{D-2}_{\rm Mink} (\epsilon) - r^{D-2}_{\rm Mink} (0) \right ) \approx  \frac{\Omega_{D-2}}{4 G \hbar} \alpha^{D-4} \epsilon^2 \; ,
\ee
which is precisely equal to (\ref{nPf-1Einstein}). Evidently we can interpret (\ref{nPf-1Einstein}) as the natural increase in the entropy of the hyperboloid in the absence of heat flux, an increase that is eliminated by considering only the reversible part of the entropy change, Eq. (\ref{revS}).

We therefore have
\be 
\Delta{S}_{\rm rev}=\frac{1}{4 G \hbar}\int_{\Sigma} d \Sigma^{a}R_{ab}(p)\xi^{b}
\ee
Now we use the fact that $\Sigma$ was constructed to be a surface of constant and uniform acceleration. We can therefore associate with it a constant and uniform temperature, Eq. (\ref{temp}). Then
we have
\be 
T\Delta{S}_{\rm rev}=\frac{1}{8\pi\alpha G}\int_{\Sigma} d \Sigma^{a}R_{ab}(p)\xi^{b} \label{TDelS}
\ee
Meanwhile, the integrated energy flux into $\Sigma$ as measured by our accelerating observers is
\be
Q = \int_\Sigma d\Sigma^a T_{ab} u^b \approx \frac{1}{\alpha} \int_\Sigma d\Sigma^a T_{ab} (p) \xi^b \; . \label{DelQ}
\ee
where the energy-momentum tensor can again be approximated to leading order by its value at $p$. Now, in thermodynamics, heat is the energy that goes into macroscopically unobservable degrees of freedom. Since the interior of the future light of $p$ is fundamentally unobservable (being causally disconnected from the exterior), we identify the integrated energy flux, Eq. (\ref{DelQ}), as heat \cite{Jacobson:1995ab}.

Clausius' theorem, $Q = T \Delta S_{\rm rev}$, then tells us to equate the integrals in (\ref{DelQ}) and (\ref{TDelS}). But note that this equality holds for all choices of $\Sigma$. For example, we could have chosen a different surface $\Sigma$ by having a different choice of $\alpha$ or by varying $\epsilon$. In particular, since the surface $\Sigma$ is capped off by constant-time slices, we can also obtain a different $\Sigma$ by performing a Lorentz boost on our Riemann normal coordinate system.
It is shown in Appendix B, that this implies that the tensors contracted with $n^a$ and $\xi^b$ in the integrands of (\ref{TDelS}) and (\ref{DelQ}) must match, up to a term that always vanishes when contracted with $n^a$ and $\xi^b$. Since $n^a \xi_a = 0$, the unknown term must be proportional to the metric. We therefore have
\be 
R_{ab}+\varphi g_{ab} =8\pi G T_{ab} \; ,
\ee
where $\varphi$ is some scalar function of spacetime. We may determine this function by demanding that the Bianchi identity hold, leading finally to Einstein's equations:
\be
R_{ab}-\frac{1}{2}Rg_{ab}+\Lambda g_{ab}=8\pi G T_{ab} \; .
\ee
Thus, gravitational equations emerge out of Clausius' theorem, $Q = \Delta S_{\rm rev}/T$, when we attribute thermodynamic properties to stretched future light cones. The cosmological constant appears as an integration constant. We have reproduced Jacobson's famous result, but using a construction based on the stretched future light cone.

It is instructive to ask why $\Delta S_{\rm rev}$ had to be positive. In fact, this follows intuitively from the way we have defined $\Sigma$ as a surface of constant acceleration, a setup that is motivated by black hole physics. Consider a sphere of observers at some radius $r$, outside some spherically symmetric body, such as a black hole. The observers stay at $r$, firing their rockets to not fall in, and are therefore all subject to the same, constant acceleration. Now suppose more matter accretes on to the source, increasing its gravitational pull. Heuristically, the observers have to move outwards in order to maintain their original acceleration. Therefore a surface of constant accelerating observers increases its area when matter falls in; this is why $\Delta S_{\rm rev}$ is positive when $Q > 0$. More precisely, explicit evaluation of $Q$ from its definition, Eq. (\ref{DelQ}), yields:
\be
Q = \frac{\Omega_{D-2}}{2} \alpha^{D-3} \epsilon^2 \left (\rho + \frac{1}{D-1} \sum_i P_i \right ) \; ,
\ee
where $\rho = -T_{tt}(p)$ and $P_i = T_{ii}(p)$. We see that $Q$ is positive when the null energy condition is obeyed. Thus our stretched future light cone has $\Delta S_{\rm rev} \geq 0$ when the null energy condition holds, analogous to the area theorem for black holes. Our stretched future light cone evidently also obeys the second law of thermodynamics.

\section{Generalized Equations of Gravity}
\noindent
In the stretched light cone formulation, this result can be extended to more general theories of gravity. Extending the thermodynamic derivation of the gravitational equations to other theories of gravity has been a long-standing challenge. Many previous attempts have been made, both for specific theories of gravity such as $f(R)$ theories, and for more general diffeomorphism-invariant theories. However, all previous attempts at general derivations have been marred by errors, or appear unphysical (or both). Four early papers, which come close, deserve special mention.

Padmanabhan \cite{Padmanabhan:2009ry} attempts to rewrite the field equations in terms of thermodynamics (rather than obtaining them from thermodynamics). The author claims, without showing any calculations, that the steps can be reversed to obtain the equations from the thermodynamics. However, he uses Killing's identity for approximate Killing vectors, without apparently realizing that it fails at the same order as the equations he would be trying to derive. Moreover, his expression for the entropy appears to depend on volume, rather than area. Parikh and Sarkar \cite{Parikh:2009qs} attempt a derivation from thermodynamics, using the Noether charge. The authors recognize that Killing's identity is invalid for approximate Killing vectors, but have no convincing justification for their use of it. They consider a rectangular spacelike patch of a (stretched) local Rindler horizon and equate the difference in area between two such patches using Stokes' theorem on a timelike surface joining them. However, that timelike surface has additional boundaries that connect the edges of the rectangles (which is easiest to visualize in (2+1)-dimensional spacetime); this contribution was missed. Brustein and Hadad \cite{Brustein:2009hy} also attempt a Noether-charge derivation from thermodynamics. The authors write some equations that do not appear correct, expressing the entropy as a volume, for example. They also appear to have used Killing's identity without realizing that it fails. In their use of Stokes' theorem, they also appear to have missed the existence of extra boundary terms. Finally, Guedens et al \cite{Guedens:2011dy} recognize both the issues (failure of Killing's identity, existence of extra boundary terms) that have tripped up previous attempts at derivations. The authors deal with the Killing's identity problem by restricting integration to a very narrow strip of the Rindler horizon plane using the observation \cite{Guedens:2012sz} that Killing's identity can be made to hold approximately near a single null generator. However, they deal with the boundary term by choosing the second surface to have the same edges as the first one, while dipping down in a nearly null test-tube shape. Although they formally succeed in obtaining the gravitational equations from the variation of a Noether charge, their derivation appears unphysical, as they themselves note. For example, even for Einstein gravity, the entropy on the looping part of the test-tube shape is no longer proportional to its area.

The success of the approach in the present work, which is based on the paper by Parikh and Sarkar \cite{Parikh:2009qs}, is directly related to our use of a stretched future light cone. Because a stretched future light cone has closed spacelike sections (spheres, which, unlike the rectangular sections of Rindler planes, have no edges), there are no extra boundary terms in Stokes' theorem. And the failure of Killing's identity is not fatal because the vast majority of problematic terms integrate to zero over a sphere; the few remaining terms can be dealt with, as shown in detail in Appendix A.

Consider then the action, $I$, of a diffeomorphism-invariant theory of gravity in $D$ dimensions of the form
\be
I=\frac{1}{16\pi G}\int d^{D}x\sqrt{-g}L\left(g^{ab},R_{abcd}\right) + I_{\rm matter} \; .
\ee
Here we have written the gravitational Lagrangian, $L$, as a function of the inverse metric $g^{ab}$ and the curvature tensor $R_{abcd}$ separately. Cast in this way, the action encompasses a wide class consisting of all diffeomorphism-invariant Lagrangian-based theories of gravity that do not involve derivatives of the Riemann tensor. We then define \cite{PaddyAseem}
\be 
P^{abcd} \equiv \frac{\partial L}{\partial R_{abcd}} \; ,
\ee
where the tensor $P^{abcd}$ can be shown to have all of the algebraic symmetries of the Riemann tensor. The gravitational equation of motion of such theories is
\be
P_{a}^{\;\;cde}R_{bcde}-2\nabla^{c}\nabla^{d}P_{acdb}-\frac{1}{2}Lg_{ab}=8\pi G T_{ab} \; . \label{eom}
\ee
In particular, for Einstein gravity, we have $L = R$, and therefore
\be 
P^{abcd}_{\rm E}=\frac{1}{2}(g^{ac}g^{bd}-g^{ad}g^{bc}) \; . \label{PEinstein}
\ee
Substituting this in (\ref{eom}), we recover Einstein's equation.
 
Our goal is to derive (\ref{eom}) from local holographic thermodynamics. Here we will see that our stretched future light cone derivation of Einstein's equations extends naturally to higher-curvature theories of gravity. Our Noetheresque approach will be based on an earlier paper by one of us \cite{Parikh:2009qs}. In that work, $\Sigma$ was a planar strip of a Rindler horizon, rather than a spherical Rindler horizon. As already mentioned, this resulted in two technical problems: (i) in Stokes' theorem, $\Delta S$ did not account for all contributions from the surface $\Sigma$ because there were also extra contributions from the edges of the strip, and (ii) the failure of Killing's identity, which does not hold for approximate symmetries, led to unwanted terms that could not be eliminated over the strip. As we have already seen, choosing a spherical Rindler horizon for $\Sigma$ resolves both these issues: since a sphere has no boundaries, the problem of extra contributions in Stokes' theorem does not arise. In addition, most of the unwanted terms arising from the failure of Killing's identity integrate to zero on a sphere. Of the remaining terms, as shown in Appendix A, the leading one precisely cancels the natural expansion of the hyperboloid, and the few remaining ones can be dealt with by redefining $\xi_a$, as in the case of Einstein gravity.

Now, information about the underlying gravitational theory is encoded within the thermodynamic formula for entropy. For Einstein gravity, the entropy is one quarter of the horizon area, but for more general theories of gravity we have to generalize the Bekenstein-Hawking entropy to something else. We will take that generalization to be the Wald entropy \cite{Wald:1993nt}. To obtain the Wald entropy, one first defines the antisymmetric Noether potential $J^{ab}$, associated with the diffeomorphism $x^{a}\to x^{a}+\xi^{a}$. For theories, that do not contain derivatives of the Riemann tensor, the Noether potential is
\be
J^{ab}=-2P^{abcd}\nabla_{c}\xi_{d}+4\xi_{d}\nabla_{c}P^{abcd} \; . \label{NoetherJ}
\ee
Then, when $\xi_{a}$ is a timelike Killing vector, the Wald entropy, $S$, associated with a stationary black hole event horizon 
is proportional to the Noether charge \cite{Wald:1993nt}:
\be
S =\frac{1}{8G \hbar}\int dS_{ab}J^{ab} \; .
\ee
Substituting (\ref{NoetherJ}) and (\ref{PEinstein}), we indeed recover the Bekenstein-Hawking entropy, Eq. (\ref{Bek}), for the case of Einstein gravity.
 
Wald's construction was designed to yield an expression for the entropy of a stationary black hole in an asymptotically flat spacetime in generalized theories of gravity. As before, we will make the nontrivial assumption of local holography, meaning that this gravitational entropy
can also be attributed locally to the future light cones of arbitrary points, and even to their timelike stretched horizons, $\Sigma$. Consider then a stretched future light cone generated by $\xi_{a}$. Analogous to (\ref{areaxi}), the Wald entropy at time $t$ is
\be
S(t) =-\frac{1}{4G \hbar}\int_{\omega(t)}dS_{ab}\left(P^{abcd}\nabla_{c}\xi_{d}-2\xi_{d}\nabla_{c}P^{abcd}\right) \; .
\ee
The total change in entropy between $t = 0$ and $t = \epsilon$ is $\Delta S_{\rm tot} = S(\epsilon) - S(0)$, or
\be
\Delta S_{\rm tot}
 =\frac{1}{4G \hbar}\int_{\Sigma}d\Sigma_{a}\nabla_{b}\left(P^{abcd}\nabla_{c}\xi_{d}-2\xi_{d}\nabla_{c}P^{abcd}\right) \; ,
\ee
where we have again invoked Stokes' theorem, Eq. (\ref{stokesthm}), for an antisymmetric tensor field. Then
\be
\Delta S_{\rm tot}=\frac{1}{4G \hbar}\int_{\Sigma}d\Sigma_{a}\left[-\nabla_{b}\left(P^{adbc}+P^{acbd}\right)\nabla_{c}\xi_{d}+P^{abcd}\nabla_{b}\nabla_{c}\xi_{d}-2\xi_{d}\nabla_{b}\nabla_{c}P^{abcd}\right] \; .
\ee
For Lovelock theories of gravity, which include Einstein gravity and Gauss-Bonnet gravity, it can be shown that $\nabla_b P^{abcd} = 0$ identically and so the first two terms vanish. For other theories of gravity, however, these terms do not generically vanish. By symmetry, only the contraction with the symmetric part of $\nabla_c \xi_d$ survives. As seen from (\ref{Killingfailure}), $\xi_{a}$ satisfies Killing's equation to $\mathcal{O}(x^{2})$, except for the $i,j$ indices, which means that the term cannot generically be discarded. Define
\be
q^a \equiv \nabla_{b}\left(P^{adbc}+P^{acbd}\right)\nabla_{c}\xi_{d}
\ee
We therefore have
\be
\Delta S_{\rm tot}=\frac{1}{4 G \hbar}\int_{\Sigma} d \Sigma_{a}\left(-q^a + P^{abcd}(R_{dcbe}\xi^{e}+f_{bcd})-2\xi_{d}\nabla_{b}\nabla_{c}P^{abcd}\right) \; , \label{TdStot}
\ee
where we have again taken into account the fact that $\xi_{a}$ does not satisfy Killing's identity, Eq. (\ref{fbcd}). This generalizes (\ref{DStot}). As shown in Appendix A, just as for the case of Einstein gravity, the unwanted term $\int_{\Sigma}d\Sigma_a P^{abcd}f_{bcd}$ can be dropped by redefining $\xi_a$ and subtracting the natural entropy increase of the hyperboloid, Eq. (\ref{revS}). In Appendix A, we show that the same redefinition of $\xi_a$ can also be used to eliminate $q^a$ for the non-Lovelock theories for which it does not identically vanish.

Defining the locally measured energy as before, Eq. (\ref{DelQ}),
\be 
Q=\int_{\Sigma}d\Sigma_{a}T^{a}_{\;e}u^{e}=\frac{1}{\alpha}\int_{\Sigma}d\Sigma_{a}T^{a}_{\;e}\xi^{e} \; ,
\ee
we see that $T\Delta S_{\rm rev}= Q$ can be written as
\be
\frac{1}{8\pi\alpha G}\int_{\Sigma} d\Sigma_{a}\left(P^{abcd}R_{dcbe}-2\nabla_{b}\nabla_{c}P^{abc}_{\; \; \; \; \; e}\right) \xi^e = \frac{1}{\alpha}\int_{\Sigma}d\Sigma_{a}T^{a}_{\;e}\xi^{e} \; . \label{Clausiusgeneral}
\ee
As shown in Appendix B, the equality of these integrals under variations of $\Sigma$ implies a stronger equality of the integrands,
\be
P_{a}^{\;cde}R_{bcde}-2\nabla^{c}\nabla^{d}P_{acdb}+ \varphi g_{ab}=8\pi G T_{ab} \; ,
\ee
where $\varphi$ is an undetermined scalar function. The requirement that the energy-momentum tensor be conserved then implies that $\varphi= - \frac{1}{2}L+\Lambda'$, where $L$ is the Lagrangian and $\Lambda'$ is an integration constant. Altogether, 
\be
P_{a}^{\;cde}R_{bcde}-2\nabla^{c}\nabla^{d}P_{acdb}-\frac{1}{2}g_{ab}L+\Lambda' g_{ab}=8\pi G T_{ab} \; ,
\ee
which we recognize as having the form of the generalized Einstein's equation for our theory of gravity, Eq. (\ref{eom}). Note, however, that the cosmological constant term does not match that in (\ref{eom}), unless the integration constant $\Lambda'$ is zero. For example, if the Lagrangian $L$ already includes a cosmological term $-2 \Lambda$, then the equation of motion derived from the action will have a term $\Lambda g_{ab}$ whereas the equation we derived from thermodynamics has a term $(\Lambda + \Lambda') g_{ab}$. This discrepancy can be traced to the fact that the Wald entropy is unaffected by the cosmological constant which does not contribute to $P_{abcd}$.

To summarize: in this paper we have defined the stretched future light cone, argued that it is natural to associate temperature and holographic entropy with it, and shown that a thermodynamic equation -- the Clausius theorem $Q = \Delta S_{\rm rev}/T$ -- directly leads to the generalized Einstein equations for all diffeomorphism-invariant theories of gravity whose Lagrangian contains no derivatives of the Riemann tensor.

\bigskip
\noindent
{\bf ACKNOWLEDGMENTS}

\noindent
We are grateful for discussions with Ted Jacobson and Sudipta Sarkar. M. P. is supported in part by John Templeton Foundation Grant No. 60253 and by the Government of India DST VAJRA Faculty Scheme VJR/2017/000117.

\newpage

\section*{Appendix A: FAILURE OF KILLING'S IDENTITY}  \label{appendixA}
\noindent
In our derivation of the gravitational equations, we made critical use of the Killing identity even though we have only an approximate Killing vector. The purpose of this appendix is to justify that step, as well as to eliminate the $\int d \Sigma_a q^a$ term in (\ref{TdStot}). We denote the failure of $\xi_a$ to satisfy Killing's identity via the tensor
\be
f_{bcd} \equiv \nabla_{b}\nabla_{c}\xi_{d}-R^{e}_{\;bcd}\xi_{e} =\frac{1}{2}\left(\nabla_{d}S_{bc}-\nabla_{c}S_{db}-\nabla_{b}S_{cd}\right)
\ee
where $S_{ab} = \nabla_{(a}\xi_{b)}$ \cite{Kothawala:2010bf}. From this we see that $f_{bdc}=-f_{bcd}$. 

In evaluating $\Delta S_{\rm tot}$, we encounter integrals of the form $\int d \Sigma_a P^{abcd} (R_{dcbe} \xi^e + f_{bcd})$, as in (\ref{TdStot}). (For Einstein gravity, $P^{abcd} = \frac{1}{2} (g^{ac} g^{bd} - g^{ad} g^{bc})$.) We would like to discard $n_{a}P^{abcd}f_{bcd}$ but retain $n_{a}P^{abcd}R^{e}_{\;bcd}\xi_{e}$. This latter quantity is, to lowest order, $\mathcal{O}(x^{2})$, since $\xi_a$ and $n_a$ are both of order $x$. Hence all terms in $f_{bcd}$ of $\mathcal{O}(x)$ and lower are problematic.

In general, $f_{bcd}$ has two types of contributions because our $\xi_a$ fails to be a Killing vector in two ways. First, $\xi_a$ generates radial boosts. These are not true isometries even of Minkowski space. This contributes a term to $f_{bcd}$ of ${\cal O}(x^{-1})$ in Riemann normal coordinates. Second, we will see that in a general curved spacetime, $\xi_a$ will have to be redefined to include quadratic and higher terms. These contribute terms to $f_{bcd}$ at ${\cal O}(1)$ and ${\cal O}(x)$. Therefore, in general, $f_{bcd}$ does not vanish at the required order.

Fortunately, we do not actually need $f_{bcd}$ to vanish, as in \cite{Guedens:2012sz,Guedens:2011dy} ; rather we require only a much weaker condition, namely that the integral of the contraction $n_a P^{abcd} f_{bcd}$ vanish to ${\cal O}(x^{2})$. We shall use several tricks to deal with nonzero terms in $f_{bcd}$. First, some terms give zero when contracted with $P^{abcd}$, because of symmetry. Second, the vast majority of terms integrate to zero over the spherical spatial sections of $\Sigma$, since the integral of any odd power of a Cartesian spatial coordinate over a sphere is zero. The remaining terms are of two types: there is the $f_{bcd}$ term of ${\cal O}(x^{-1})$ that exists even in Minkowski space,
and there are a small handful of leftover $f_{bcd}$ terms of ${\cal O}(1)$ and ${\cal O}(x)$ in curved space. The integral of the first term does not vanish. However, as we show, it is precisely canceled by subtracting the component of $T\Delta S$ that comes from the natural expansion of $\Sigma$. The other terms can be eliminated by redefining the higher-order terms in $\xi_a$, as we will show.

Our integrand $\sqrt{g} n_a P^{abcd} f_{bcd}$ will have various order pieces ranging from $\mathcal{O}(1)$ to $\mathcal{O}(x^{2})$, with higher orders negligible. We need to show that the integral at each order either vanishes or can be canceled. Let us first classify each of the terms. We do this by expanding 
 \be
n_{a}\approx n^{(1)}_{a}+n^{(2)}_{a}+n^{(3)}_{a},\quad P^{abcd}\approx P^{abcd}_{(0)}+P^{abcd}_{(1)}+P^{abcd}_{(2)},\quad f_{bcd}\approx f^{{\cal O}(-1)}_{bcd}+f^{(0)}_{bcd}+f^{(1)}_{bcd}
 \ee
where the subscript or superscript indicates the order, in $x$, of the given quantity. We also note that for the integration measure we have $\sqrt{g}\approx\sqrt{\eta}+\sqrt{h}$ which is of $\mathcal{O}(1)+\mathcal{O}(x^{2})$. 

Then the lowest order contribution to the offending term is 
\be
  \frac{1}{4G \hbar}\int_{\Sigma}dA d \tau n^{(1)}_{a}P^{abcd}_{(0)}f^{{\cal O}(-1)}_{bcd}\label{npf0}
\ee
which is of $\mathcal{O}(1)$. The next order terms, of $\mathcal{O}(x)$, are given by
 \be
  \frac{1}{4G \hbar}\int_{\Sigma}dAd \tau \left(n^{(1)}_{a}P^{abcd}_{(1)}f^{{\cal O}(-1)}_{bcd}+n^{(2)}_{a}P^{abcd}_{(0)}f^{{\cal O}(-1)}_{bcd}+n^{(1)}_{a}P^{abcd}_{(0)}f^{(0)}_{bcd}\right)\label{npf1}
   \ee
Last, the highest order term we need consider is
 \be
 \begin{split}
\frac{1}{4G \hbar} \int_{\Sigma}dA d \tau&\biggr\{\sqrt{h}n^{(1)}_{a}P^{abcd}_{(0)}f^{{\cal O}(-1)}_{bcd}+n^{(1)}_{a}P^{abcd}_{(2)}f^{{\cal O}(-1)}_{bcd}+n^{(1)}_{a}P^{abcd}_{(1)}f^{(0)}_{bcd}+n^{(1)}_{a}P^{abcd}_{(0)}f^{(1)}_{bcd}\\
 &+n^{(2)}_{a}P^{abcd}_{(1)}f^{{\cal O}(-1)}_{bcd}+n^{(2)}_{a}P^{abcd}_{(0)}f^{(0)}_{bcd}+n^{(3)}_{a}P^{abcd}_{(0)}f^{{\cal O}(-1)}_{bcd}\biggr\}
 \end{split}
\label{npf2} 
\ee
which is clearly of $\mathcal{O}(x^{2})$. We therefore need to show (\ref{npf0}), (\ref{npf1}), and (\ref{npf2}) vanish for an arbitrary $P^{abcd}$. Let us begin with (\ref{npf0}).

\subsection*{Removing the Natural Expansion of the Hyperboloid}
\indent
Writing out $f_{bcd}$ explicitly, we have
\be
f_{bcd}=\partial_{b}\partial_{c}\xi_{d}+\left(2\Gamma^{f}_{\;b(c}\Gamma^{e}_{\;d)f}-\partial_{b}\Gamma^{e}_{\;cd}\right)\xi_{e}-\left(\Gamma^{e}_{\;bc}\partial_{e}\xi_{d}+2\Gamma^{e}_{\;d(c}\partial_{b)}\xi_{e}\right)-R^{e}_{\;bcd}\xi_{e} \label{fbcd-explicit}  
\ee
Note that $\xi_{a}$, $n_a$, and the Christoffel symbols are all of $\mathcal{O}(x)$. 
Therefore the term $n_{a}2\Gamma^{f}_{\;b(c}\Gamma^{e}_{\;d)f}\xi_{e}$ is of much higher order than the rest of the terms and we can neglect it. Moreover, given that $P^{abcd}$ is antisymmetric in its final two indices and $\Gamma^{e}_{\;cd,b}$ is symmetric in $c$ and $d$, it will not contribute to $n_a P^{abcd} f_{bcd}$. Therefore, we need only consider the reduced expression:
\be
f_{bcd} \approx \partial_{b}\partial_{c}\xi_{d}-2\Gamma^{e}_{\;bc}\partial_{[e}\xi_{d]}-R^{e}_{\;bcd}\xi_{e}
\ee
 
To lowest order, we have
\be
f_{bcd}^{\mathcal{O}(-1)}=\partial_{b}\partial_{c}\xi^{\mathcal{O}(1)}_{d}
\ee
From (\ref{Killingfailure}), we find that Killing's identity, at $\mathcal{O}(x^{-1})$, fails as,
 \be
 \begin{split}
 &f_{tij}^{\mathcal{O}(-1)}=f_{itj}^{\mathcal{O}(-1)}=-f_{ijt}^{\mathcal{O}(-1)}=\frac{1}{r}\left(\delta_{ij}-\frac{x_{i}x_{j}}{r^{2}}\right)\\
& f_{ijk}^{\mathcal{O}(-1)}=-\frac{t}{r^{3}}\left(x_{i}\delta_{jk}+x_{j}\delta_{ik}+x_{k}\delta_{ij}\right)+\frac{3t}{r^{5}}x_{i}x_{j}x_{k}
 \label{fO-1bcd}\end{split}
 \ee
Using the algebraic symmetries of $P^{abcd}$ and $f^{\mathcal{O}(-1)}_{bcd}$, we have
\be
P^{abcd}f^{\mathcal{O}(-1)}_{bcd}=P^{aijk}f_{ijk}^{\mathcal{O}(-1)}+P^{atij}f^{\mathcal{O}(-1)}_{tij}+P^{aitj}f_{itj}^{\mathcal{O}(-1)}+P^{aijt}f_{ijt}^{\mathcal{O}(-1)} =2P^{aitj}f_{itj}^{\mathcal{O}(-1)}
\ee
The undesired term then becomes
\be
\begin{split}
 \frac{1}{4G\hbar}\int_{\Sigma}dAd\tau n_{a}P^{abcd}f_{bcd}^{\mathcal{O}(-1)}&=\frac{1}{4G\hbar}\int_{\Sigma}dAd\tau\left(2n_{t}P^{titj}f^{\mathcal{O}(-1)}_{itj}+2n_{i}P^{tkij}f^{\mathcal{O}(-1)}_{jtk}\right)\\
 &=-\frac{1}{4G\hbar}\int_{\Sigma}dAd\tau\frac{2t}{\alpha r}P^{titj}\left(\delta_{ij}-\frac{x_{i}x_{j}}{r^{2}}\right)
 \end{split}
 \ee
 where in the last step we used spherical symmetry killing off all integrals with parity. Moreover, by parity, this term will vanish for all terms $i\neq j$, keeping only terms with $i=j$. With this fact in mind, and using that $d\tau = dt \alpha/r$, and $\sum x_{i}^{2}=r^{2}$, we have
 \be
\begin{split}
 \frac{1}{4G\hbar}\int_{\Sigma}dAd\tau n_{a}P^{abcd}f_{bcd}^{\mathcal{O}(-1)}&=-\frac{1}{4G\hbar}(D-2)\frac{2\sum_{i}P^{titi}}{\alpha (D-1)}\left(\int d\Omega_{D-2}\right)\int_{0}^{t_{0}}dt\frac{\alpha}{r} r^{D-3}t\\
 &=-\frac{1}{2(D-1)G\hbar}(D-2)\sum_{i}P^{titi}\Omega_{D-2}\int_{0}^{t_{0}}dt\left(\alpha^{2}+t^{2}\right)^{(D-4)/2}t\\
 &=-\frac{1}{2(D-1)G\hbar}\sum_{i}P^{titi}\Omega_{D-2}\left[\left(\alpha^{2}+t^{2}_{0}\right)^{(D-2)/2}-\alpha^{(D-2)}\right] \label{leadingnpf}
 \end{split}
 \ee

Recall that we are applying Clausius' theorem, $T\Delta S_{\rm rev}= Q$, to derive the equations of motion for an arbitrary theory of gravity. But $\Delta S_{\rm tot}$ includes all change in the entropy, not just the change in entropy due to the heat flow through $\Sigma$. In particular, even in the absence of heat flow, the entropy increases because of the natural increase in the area of a congruence of outwardly accelerating observers.

Let us calculate the increase in entropy from the natural background expansion of the hyperboloid. Begin with the Wald entropy,
 \be
S =\frac{1}{8G\hbar}\int_{S}dS_{ab}J^{ab}=-\frac{1}{4G\hbar}\int_{S}dS_{ab}\left(P^{abcd}\nabla_{c}\xi_{d}-2\xi_{d}\nabla_{c}P^{abcd}\right)\;.\label{SW}
  \ee
To leading order we can neglect the $\nabla_{c}P^{abcd}$ term. Substituting in our leading-order expressions for the outward pointing normal $n_{a}$, and $u_{a}=\xi_{a}/\alpha$, we find
 \be 
 \begin{split}
S &=-\frac{1}{4G\hbar}\int_{S}dA\left(n_{t}u_{i}-n_{i}u_{t}\right)\left[P^{titj}2\partial_{t}\xi_{j}+P^{tijk}\partial_{j}\xi_{k}\right]\\
&=-\frac{1}{4G\hbar}\int_{S}dA\frac{x_{i}}{r}\left[2P^{titj}\partial_{t}\xi_{j}+P^{tijk}\partial_{j}\xi_{k}\right]\\
&=-\frac{1}{4G\hbar}\int_{S}dA\left(2P^{titj}\frac{x_{i}x_{j}}{r^{2}}\right)\\
&=-\frac{1}{2(D-1)G\hbar}\sum_{i}P^{titi}\Omega_{D-2}r^{D-2}(t_{0})\;,
 \end{split}
 \ee
where we used parity to move to the final line. We are interested in the change in entropy, $\Delta S_{\rm hyp}$, due to the expansion of the hyperboloid. Using $r_{\rm hyp}(t) = (\alpha^2 + t^2)^{1/2}$, we find
\be
\begin{split}
\Delta S_{\rm hyp} &\equiv S_{\rm hyp}(t_{0})-S_{\rm hyp}(0)=-\frac{1}{2(D-1)G\hbar}\sum_{i}P^{titi}\Omega_{D-2}\left[r_{\rm hyp}^{D-2}(t_{0})-r_{\rm hyp}^{D-2}(0)\right]\\
&=-\frac{1}{2(D-1)G\hbar}\sum_{i}P^{titi}\Omega_{D-2}\left[(\alpha^{2}+t_{0}^{2})^{(D-2)/2}-\alpha^{(D-2)}\right]\;,
\end{split}
\ee
which precisely matches the leading-order part of the term, Eq. (\ref{leadingnpf}), we are trying to eliminate:
 \be 
  \Delta S_{\rm hyp}= \frac{1}{4G\hbar}\int_{\Sigma}dA d \tau n_{a}P^{abcd}f_{bcd}^{\mathcal{O}(-1)}\;.
  \ee
That is, the unwanted term is exactly equal to the entropy due to the natural expansion of the hyperboloid. This term should be subtracted from $\Delta S_{\rm tot}$ before equating it to $Q$. Moreover, note that here we did not specify the exact form of $P^{abcd}$, and therefore this subtraction holds for arbitrary theories of gravity. 

\subsection*{Eliminating Higher Order Contributions}

Now we must deal with the higher order contributions, namely $\mathcal{O}(x)$ and $\mathcal{O}(x^{2})$. As alluded to above, in order to eliminate the higher order contributions to $n_{a}P^{abcd}f_{bcd}$, we consider a more generic $\xi_{a}$ and $n_{a}$, namely, 
\be
\begin{split}
\xi_{a}&=\xi^{(1)}_{a}+\xi^{(2)}_{a}+\xi^{(3)}_{a}+...\\
&=-r\delta_{ta}+\frac{tx^{i}}{r}\delta_{ia}+\frac{1}{2!}C_{\mu\nu a}x^{\mu}x^{\nu}+\tilde{C}_{\nu a}rx^{\nu}+\frac{1}{3!}D_{\mu\nu \rho a}x^{\mu}x^{\nu}x^{\rho}+\frac{1}{2!}\tilde{D}_{\mu\nu a}rx^{\mu}x^{\nu}+...
\end{split}
\ee
\be
\begin{split}
\alpha n_{a}&=\alpha(n^{(1)}_{a}+n^{(2)}_{a}+n^{(3)}_{a}+...)\\
&=-t\delta_{at}+x^{i}\delta_{ai}+\frac{1}{2!}C'_{\mu\nu a}x^{\mu}x^{\nu}+\frac{1}{3!}D'_{\mu\nu\rho a}x^{\mu}x^{\nu}x^{\rho}+...
\end{split}
\ee
Here we adopt the notation that $\mu,\nu,\rho...,$ represent the full spacetime index while $i,j,k,\ell,h$ represent spatial components, and where $\xi_{a}^{(\cdot)}$ denotes the order of the component; e.g., $\xi^{(1)}_{a}=-r\delta_{ta}+\frac{tx^{i}}{r}\delta_{ia}$ is of order $\mathcal{O}(x)$. 

Let us substitute our modified $\xi_{a}$ into our expression for $f_{bcd}$, for which we reproduce the simplified version here for convenience:
 \be
  f_{bcd}=\partial_{b}\partial_{c}\xi_{d}-\Gamma^{e}_{\;bc}\partial_{e}\xi_{d}-R^{e}_{\;bcd}\xi_{e}\;.\label{fbcdred}
  \ee
We have already worked out the $f^{\mathcal{O}(-1)}_{bcd}$ terms (\ref{fO-1bcd}).
 
 Next, the only possible term in $f_{bcd}$ of order $\mathcal{O}(1)$ is
 \be
  f^{\mathcal{O}(0)}_{bcd}\equiv\partial_{b}\partial_{c}\xi^{(2)}_{d}=C_{bcd}\;.
  \ee
Now let us work out the term in $f_{bcd}$ of order $\mathcal{O}(x)$. This will include a combination of terms including $\partial_{b}\partial_{c}\xi_{d}^{\mathcal{O}(3)}$, and the remaining terms in (\ref{fbcdred}) of order $\mathcal{O}(x)$, namely,
 \be
 \partial_{b}\partial_{c}\xi_{d}^{(3)}=D_{\nu bcd}x^{\nu}+r\tilde{D}_{bcd}+\tilde{D}_{\nu cd}(\partial_{b}r)x^{\nu}+\tilde{D}_{\nu bd}(\partial_{c}r)x^{\nu}+\frac{1}{2!}\tilde{D}_{\mu\nu d}x^{\mu}x^{\nu}(\partial_{b}\partial_{c}r)
 \ee
 \be
 -2\Gamma^{e}_{\;bc}(h)\partial_{[e}\xi^{\mathcal{O}(1)}_{d]}+\mathcal{O}(x^{2})
 \ee
 \be 
  R^{e}_{\;bcd}(p)\xi^{(1)}_{e}+\mathcal{O}(x^{2})\;,
  \ee
 where 
 \be 
  \Gamma^{e}_{\;bc}(h)\equiv\frac{1}{2}\eta^{ef}\left(\partial_{b}h_{cf}+\partial_{c}h_{bf}-\partial_{f}h_{bc}\right)
=-\frac{x^{\mu}}{3}\eta^{ef}(R_{c\mu fb}+R_{b\mu fc})\;,
  \ee
and we used $h_{ab}=-\frac{1}{3} R_{a\mu b\nu}x^{\mu}x^{\nu}$. Moreover,  since
  \be 
  \partial_{i}\xi_{t}^{\mathcal{O}(1)}=-\frac{x_{i}}{r}=-\partial_{t}\xi^{\mathcal{O}(1)}_{i}\;,
 \ee
  the only nonvanishing contribution to $\partial_{[e}\xi_{d]}$ is $\partial_{[i}\xi_{t]}=-\frac{x_{i}}{r}$.
  Altogether, one finds:
  \be
  \begin{split}
  f_{bcd}^{\mathcal{O}(1)}&=\partial_{b}\partial_{c}\xi^{\mathcal{O}(3)}_{d}-2\Gamma^{e}_{\;bc}(h)\partial_{[e}\xi^{\mathcal{O}(1)}_{d]}-R^{e}_{\;bcd}\xi_{e}^{\mathcal{O}(1)}\;.
  \end{split}
\ee
Note that this is the highest order of $f_{bcd}$ we need to keep since any higher order would give at least an $\mathcal{O}(x^{3})$ contribution to the integrand of the offending term, which we neglect. 
 
Recall that we need to eliminate (\ref{npf0}), (\ref{npf1}), and (\ref{npf2}) for an arbitrary $P^{abcd}$. We have already dealt with  (\ref{npf0}).
Before we go through the minutiae of these calculations, let us first explain the aim of the next two subsections providing us with a tether to hold onto as we work through the details.
 
The general prescription in eliminating the higher order contributions to $n_{a}P^{abcd}f_{bcd}$ is as follows. The integrand will include all sorts of monomial contributions, e.g., $t^{3}x_{i}x_{j}/r^{3}$. Since we care about the integral $\int_{\Sigma}n_{a}P^{abcd}f_{bcd}$ vanishing -- not the integrand -- we see that several of the monomials do not end up contributing to the final result; for example, $t^{3}x_{i}x_{j}/r^{3}$ will vanish for all $i\neq j$ as we are integrating over a sphere. Therefore we need only concern ourselves with, e.g., $t^{3}(x_{i})^{2}/r^{3}$. 
 
While these greatly reduce the number of monomial contributions, we still cannot fully eliminate the entire $\int_{\Sigma}n_{a}P^{abcd}f_{bcd}$. This is why we modify $\xi_{a}$ and $n_{a}$. More specifically, there are only a select few combinations of monomials which will appear in the integrand that do not vanish upon integration over the sphere. By modifying $\xi_{a}$ and $n_{a}$ we do not change the number of monomial contributions. Instead we find our modifications to $\xi_{a}$ and $n_{a}$ give us sets of coefficients that allow us the freedom to eliminate all other monomials, provided we have enough coefficients to do so. In short, we have a counting argument: If the number of nonvanishing monomials is less than the number of coefficients contributing to the same monomial, we can potentially force each monomial contribution to zero, i.e., $\int_{\Sigma}n_{a}P^{abcd}f_{bcd}\to0$ with a judicious choice of coefficients. 
 
In what follows we use this general prescription to separately eliminate monomials of order $\mathcal{O}(x)$ and $\mathcal{O}(x^{2})$. With the benefit of hindsight, we realize that only certain modifications to $\xi_{a}$ and $n_{a}$ will aid us, particularly,
\be
\begin{split}
\xi_{a}&=\xi^{(1)}_{a}+\xi^{(2)}_{a}+\xi^{(3)}_{a}+...\\
&=-r\delta_{ta}+\frac{tx^{i}}{r}\delta_{ia}+\tilde{C}_{\nu a}rx^{\nu}+\frac{1}{3!}D_{\mu\nu\rho a}x^{\mu}x^{\nu}x^{\rho}\;,
\end{split}
\ee
\be
\begin{split}
\alpha n_{a}&=\alpha(n^{(1)}_{a}+n^{(3)}_{a}+...)\\
&=-t\delta_{at}+x^{i}\delta_{ai}+\frac{1}{3!}D'_{\mu\nu\rho a}x^{\mu}x^{\nu}x^{\rho}\;.
\end{split}
\ee
As we will now explicitly show, this will be enough to cancel all undesired contributions coming from $\int_{\Sigma}n_{a}P^{abcd}f_{bcd}$ through $\mathcal{O}(x^{2})$. (Note that although we have set $n_{a}^{(2)}$ to zero, if we insist that $n_{a}$ be orthogonal to $\xi_{a}$ at order $\mathcal{O}(x^{3})$, we should include an $n^{(2)}_{a}$ contribution of the form $\tilde{C}'_{\nu a}tx^{\nu}$. It can be tediously verified that adding such terms to $n_{a}$ does not affect the counting argument, allowing us to leave them off in what follows.)
        
\subsubsection*{$\mathcal{O}(x)$ Contributions}
 
With the $n_{a}^{\mathcal{O}(2)}$ term being set to zero, the $\mathcal{O}(x)$ term to be eliminated becomes
\be
\frac{1}{4}\int_{\Sigma}dA d \tau\left(n^{\mathcal{O}(1)}_{a}P^{abcd}_{\mathcal{O}(1)}f^{\mathcal{O}(-1)}_{bcd}+n^{\mathcal{O}(1)}_{a}P^{abcd}_{\mathcal{O}(0)}f^{\mathcal{O}(0)}_{bcd}\right)\;.
\ee
Let us first list the various types of monomial contributions which might appear in the integrand:
\be
\mathcal{O}(x):\quad t,\;r,\;\frac{(x_{i})^{2}}{r},\;\frac{t^{2}(x_{i})^{2}}{r^{3}},\;\frac{(x_{i})^{2}(x_{j})^{2}}{r^{3}},\;\frac{(x_{i})^{4}}{r^{3}}\;.
\label{monoO1}\ee
As we will verify explicitly in a moment, only a subset of these monomials appear. Following the outlined prescription above, we need to check that we have enough coefficients to remove each of the monomial contributions. The only coefficients which will appear are those coming from the $f_{bcd}^{\mathcal{O}(0)}$ contribution, specifically $\tilde{C}_{na}$, for which we have $D^{2}$ coefficients. The number of problematic monomials which might appear is $1+1+1+(D-2)+(D-2)+\frac{1}{2}(D-1)(D-2)=D(D+1)/2<D^{2}$, for $D\geq3$. Therefore it already seems plausible that we will in fact have far more than enough coefficients to eliminate all of the monomial contributions appearing in the integrand.  Let us now verify this in detail. 

As was worked out in the previous section, we have
\be
 P^{abcd}f_{bcd}^{\mathcal{O}(-1)}=2P^{aitj}f^{\mathcal{O}(-1)}_{itj}=\frac{2}{r}P^{aitj}\left(\delta_{ij}-\frac{x_{i}x_{j}}{r^{2}}\right)\;.
\ee
Hence
\be
\begin{split}
n^{\mathcal{O}(1)}_{a}P^{abcd}_{\mathcal{O}(1)}f^{\mathcal{O}(-1)}_{bcd}&=\frac{2}{r}\left(\delta_{ij}-\frac{x_{i}x_{j}}{r^{2}}\right)\left[-\frac{t}{\alpha}P^{titj}_{\mathcal{O}(1)}+\frac{x_{k}}{\alpha}P^{kitj}_{\mathcal{O}(1)}\right]\\
&=\frac{2}{\alpha r}x_{k}\delta_{ij}P^{kitj}_{\mathcal{O}(1)}-\frac{2t}{\alpha r}\left(\delta_{ij}-\frac{x_{i}x_{j}}{r^{2}}\right)P^{titj}_{\mathcal{O}(1)}\;.
\end{split}
\ee
Defining
\be
P^{titj}_{\mathcal{O}(1)}\equiv\mathcal{P}_{\mathcal{O}(1),\mu}^{titj}x^{\mu}\quad P^{kitj}_{\mathcal{O}(1)}=\mathcal{P}^{kitj}_{\mathcal{O}(1),\mu}x^{\mu}\;,
\ee
we find that the only contributing terms to the integrand, i.e., those which do not vanish via parity arguments, are
\be
\begin{split}
n^{\mathcal{O}(1)}_{a}P^{abcd}_{\mathcal{O}(1)}f^{\mathcal{O}(-1)}_{bcd}&=-\frac{2}{\alpha r}\left(\delta_{ij}-\frac{x_{i}x_{j}}{r^{2}}\right)t^{2}\mathcal{P}^{titj}_{\mathcal{O}(1),t}+\frac{2}{\alpha r}\delta_{ij}x_{k}x^{\ell}\mathcal{P}^{kitj}_{\mathcal{O}(1),\ell}\;,
\end{split}
\ee
where we have used $x_{k}x_{i}P^{ikcd}=0$ using the symmetries of $P^{abcd}$. 

Generally, then, we see that only certain monomials appear which need to be removed. Specifically, 
\be
\begin{split}
n^{\mathcal{O}(1)}_{a}P^{abcd}_{\mathcal{O}(1)}f^{\mathcal{O}(-1)}_{bcd}&=\frac{A}{\alpha}\frac{t^{2}}{r}+\frac{A^{ii}}{\alpha}\frac{t^{2}(x_{i})^{2}}{r^{3}}+\frac{B^{ii}}{\alpha}\frac{(x_{i})^{2}}{r}\;,
\end{split}
\ee
where we have defined
\be
A\equiv-2\delta_{ij}\mathcal{P}^{titj}_{\mathcal{O}(1),t}\,,\quad A^{ii}\equiv2\mathcal{P}^{titi}_{\mathcal{O}(1),t}\,,\quad B^{k}_{\;\ell}\equiv 2\delta_{ij}\mathcal{P}^{kitj}_{\mathcal{O}(1),\ell}\;.
\ee
We now show that modifying $\xi_{a}$ via 
\be
\xi_{a}^{\mathcal{O}(2)}= r\tilde{C}_{\mu a}x^{\mu}
 \ee
will eliminate all the above undesired contributions. We have
\be 
\begin{split}
\partial_{b}\partial_{c}\xi^{\mathcal{O}(2)}_{d}&=\partial_{b}\left[\tilde{C}_{\mu d}(\partial_{c}r)x^{\mu}+\tilde{C}_{cd}r\right]\\
&=\tilde{C}_{\mu d}(\partial_{b}\partial_{c}r)x^{\mu}+\tilde{C}_{bd}(\partial_{c}r)+\tilde{C}_{cd}(\partial_{b}r)\;.
\end{split}
\ee
Then, using
\be
\partial_{i}r=\frac{x_{i}}{r}\,,\quad \partial_{i}\partial_{j}=\frac{1}{r}\left(\delta_{ij}-\frac{x_{i}x_{j}}{r^{2}}\right)\;,
\ee
we find
\be
\partial_{i}\partial_{j}\xi^{\mathcal{O}(2)}_{d}=\tilde{C}_{\mu d}\frac{x^{\mu}}{r}\left(\delta_{ij}-\frac{x_{i}x_{j}}{r^{2}}\right)+\tilde{C}_{id}\frac{x_{j}}{r}+\tilde{C}_{jd}\frac{x_{i}}{r}\;,
\ee
\be
\partial_{i}\partial_{t}\xi^{\mathcal{O}(2)}_{d}=\tilde{C}_{td}\frac{x_{i}}{r}\,,\quad \partial_{t}^{2}\xi^{\mathcal{O}(2)}_{d}=0\;.
\ee
Using these relations we find that
\be
\begin{split}
n^{\mathcal{O}(1)}_{a}P^{abcd}_{\mathcal{O}(0)}f_{bcd}^{\mathcal{O}(0)}&=\frac{1}{\alpha}\biggr\{-tP^{titj}_{\mathcal{O}(0)}(\partial_{t}\partial_{t}\xi^{\mathcal{O}(2)}_{j})-tP^{tijk}_{\mathcal{O}(0)}(\partial_{i}\partial_{j}\xi^{\mathcal{O}(2)}_{k})-tP^{tijt}_{\mathcal{O}(0)}(\partial_{i}\partial_{j}\xi^{\mathcal{O}(2)}_{t})\\
&+x_{i}P^{ijtk}_{\mathcal{O}(0)}(\partial_{j}\partial_{i}\xi_{k}^{\mathcal{O}(2)})+x_{i}P^{ijk\ell}_{\mathcal{O}(0)}(\partial_{j}\partial_{k}\xi^{\mathcal{O}(2)}_{\ell})+x_{i}P^{ijkt}_{\mathcal{O}(0)}(\partial_{i}\partial_{j}\xi_{t}^{\mathcal{O}(2)})\biggr\}\\
&=\frac{1}{\alpha r}\biggr\{-t^{2}\left(\delta_{ij}-\frac{x_{i}x_{j}}{r^{2}}\right)\left[\tilde{C}_{tk}P^{tijk}_{\mathcal{O}(0)}+\tilde{C}_{tt}P^{tijt}_{\mathcal{O}(0)}\right]\\
&+\left[\tilde{C}_{h\ell}P^{ijk\ell}_{\mathcal{O}(0)}+\tilde{C}_{ht}P^{ijkt}_{\mathcal{O}(0)}\right]\delta_{jk}x_{i}x^{h}+\left[\tilde{C}_{j\ell}P^{ijk\ell}_{\mathcal{O}(0)}+\tilde{C}_{jt}P^{ijkt}_{\mathcal{O}(0)}\right]x_{k}x_{i}\biggr\}\;.
\end{split}
\ee

Combining this with the term we wish to eliminate gives
\be
\left[\frac{A}{\alpha}-\frac{\delta_{ij}}{\alpha}(P^{tijt}_{\mathcal{O}(0)}\tilde{C}_{tt}+\tilde{C}_{tk}P^{tijk}_{\mathcal{O}(0)})\right]\frac{t^{2}}{r}
\ee
and
\be
\left[\frac{A^{ii}}{\alpha}+\frac{1}{\alpha}(\tilde{C}_{tt}P^{tiit}_{\mathcal{O}(0)}+\tilde{C}_{tk}P^{tiik}_{\mathcal{O}(0)})\right]\frac{t^{2}}{r^{3}}(x_{i})^{2}\;,
\ee
and last, 
\be
\left[\frac{B^{ii}}{\alpha}+\frac{1}{\alpha}(\tilde{C}^{i}_{\;\ell}P^{ijk\ell}_{\mathcal{O}(0)}+\tilde{C}^{i}_{\;t}P^{ijkt}_{\mathcal{O}(0)})\delta_{jk}+\frac{1}{\alpha}(\tilde{C}_{j\ell}P^{iji\ell}_{\mathcal{O}(0)}+\tilde{C}_{jt}P^{ijit}_{\mathcal{O}(0)})\right]\frac{(x_{i})^{2}}{r}\;.
\ee
The first two of these gives us $1+(D-2)=(D-1)$ monomials to cancel. But to remove these monomials, we have $1+(D-1)=D$ coefficients to work with, giving us  enough coefficients to cancel all of the undesired terms. Studying the problem at this level has provided us with insight that will prove useful when we study the elimination of $\mathcal{O}(x^{2})$ terms: (i) Not all of the possible monomials appear, and (ii) not all of the possible coefficients we have to work with will appear. Despite this we will still have enough coefficients to achieve our goal of removing $\int_{\Sigma}n_{a}P^{abcd}f_{bcd}$. 

\subsubsection*{$(2+1)$-Dimensional $f(R)$-gravity: A Restrictive Case}

Based on the above calculation, however, it is clear that if one of the quantities multiplying a set of the coefficients vanishes, e.g., $P^{tijk}$, then we might be in trouble as we can no longer use these coefficients. This is precisely the case for $f(R)$ theories of gravity (except Einstein gravity, for which there is no $P^{abcd}_{\mathcal{O}(1)}$ contribution to be canceled and we can set all $\tilde{C}$ coefficients to zero). Thus, the most restrictive case is $(2+1)$-dimensional $f(R)$ gravity. Let us study this particular example explicitly and verify that we still have enough coefficients to eliminate all monomials. 

In $f(R)$ gravity one has
\be
P^{abcd}_{f(R)}=\frac{f'(R)}{2}(g^{ac}g^{bd}-g^{ad}g^{bc})\;.
\ee
So, 
\be
P^{abcd}_{f(R),\mathcal{O}(0)}=\frac{f'(R)(p)}{2}(\eta^{ac}\eta^{bd}-\eta^{ad}\eta^{bc})\,,\;\; P^{abcd}_{f(R),\mathcal{O(1)}}=\frac{f'(R)(x)}{2}(\eta^{ac}\eta^{bd}-\eta^{ad}\eta^{bc})\equiv \mathcal{P}^{abcd}_{\mathcal{O}(1),\mu}x^{\mu}\;,
\ee
where $p$ is the spacetime point where these expressions are being evaluated. This tells us that $B^{ii}=0$, leaving
\be
\left[\frac{A}{\alpha}-\frac{\delta_{ij}}{\alpha}P^{tijt}_{\mathcal{O}(0)}\tilde{C}_{tt}\right]\frac{t^{2}}{r}
\ee
and
\be
\left[\frac{A^{ii}}{\alpha}+\frac{1}{\alpha}\tilde{C}_{tt}P^{tiit}_{\mathcal{O}(0)}\right]\frac{t^{2}}{r^{3}}(x_{i})^{2}\;,
\ee
where
\be
A=-2\delta_{ij}\mathcal{P}^{titj}_{\mathcal{O}(1),t}\,,\quad A^{ii}=\mathcal{P}^{titi}_{\mathcal{O}(1),t}\;.
\ee
Expanding our above expressions in a $(2+1)$-dimensional spacetime yields
\be
\frac{1}{\alpha}\left[-2(\mathcal{P}^{txtx}_{\mathcal{O}(1),t}+\mathcal{P}^{tyty}_{\mathcal{O}(1),t})+\tilde{C}_{tt}(P^{txtx}_{\mathcal{O}(0)}+P^{tyty}_{\mathcal{O}(0)})\right]\frac{t^{2}}{r}
\ee
and
\be
\frac{1}{\alpha}\left[2(\mathcal{P}^{txtx}_{\mathcal{O}(1),t}x^{2}+\mathcal{P}^{tyty}_{\mathcal{O}(1),t}y^{2})-\tilde{C}_{tt}(P^{txtx}_{\mathcal{O}(0)}x^{2}+P^{tyty}_{\mathcal{O}(0)}y^{2})\right]\frac{t^{2}}{r^{3}}
\ee

Each of these must vanish separately. Using that 
\be
 P^{txtx}_{\mathcal{O}(0)}=P^{tyty}_{\mathcal{O}(0)}\,,\quad \mathcal{P}^{txtx}_{\mathcal{O}(1),t}=\mathcal{P}^{tyty}_{\mathcal{O}(1),t}\;,
 \ee
 we are led to
 \be
 \frac{1}{\alpha}\left(-4\mathcal{P}^{titi}_{\mathcal{O}(1),t}+2\tilde{C}_{tt}P^{titi}_{\mathcal{O}(0)}\right)\frac{t^{2}}{r}\;,	 \label{measurezeroex}
 \ee
 \be
 \frac{1}{\alpha}\left(2\mathcal{P}^{titi}_{\mathcal{O}(1),t}-\tilde{C}_{tt}P^{titi}_{\mathcal{O}(0)}\right)\frac{t^{2}(x^{2}+y^{2})}{r^{3}}\;.
 \ee
Since $x^{2}+y^{2}=r^{2}$, we find that the above two conditions are in fact the same; miraculously the monomials add in such a way that we need only a single coefficient. (In fact, this feature of two seemingly different conditions becoming one can readily be obtained in this case if one uses the fact that  $P^{titj}_{\mathcal{O}(0)}\left(\delta_{ij}-\frac{x_{i}x_{j}}{r^{2}}\right)=-\frac{f'(R)(p)}{2}(D-2)$ from the start.) Finally, it is possible in principle that, say, $P^{titi}_{\mathcal{O}(0)}$ vanishes while $\mathcal{P}^{titi}_{\mathcal{O}(1),t}$ does not, preventing (\ref{measurezeroex}) from being set to zero. However, inspecting (\ref{measurezeroex}), it is easy to see that this can happen at most on a set of measure zero.
 
\subsubsection*{$\mathcal{O}(x^{2})$ Contributions}

Let us now move on to the $\mathcal{O}(x^{2})$ contribution to $n_{a}P^{abcd}f_{bcd}$ where the story and prescription are the same, though far more tedious to work out. Setting $n_{a}^{\mathcal{O}(2)}$ to zero means that we must eliminate
 \be
 \begin{split}
\frac{1}{4} \int_{\Sigma}dA d \tau&\biggr\{\sqrt{h}n^{\mathcal{O}(1)}_{a}P^{abcd}_{\mathcal{O}(0)}f^{\mathcal{O}(-1)}_{bcd}+n^{\mathcal{O}(1)}_{a}P^{abcd}_{\mathcal{O}(2)}f^{\mathcal{O}(-1)}_{bcd}+n^{\mathcal{O}(1)}_{a}P^{abcd}_{\mathcal{O}(1)}f^{\mathcal{O}(0)}_{bcd}+n^{\mathcal{O}(1)}_{a}P^{abcd}_{\mathcal{O}(0)}f^{(1)}_{bcd}\\
 &+n^{\mathcal{O}(3)}_{a}P^{abcd}_{\mathcal{O}(0)}f^{\mathcal{O}(-1)}_{bcd}\biggr\}\;.
 \end{split}
\ee
At the $\mathcal{O}(x^{2})$ level, the only monomials which might appear are
\be
 t^{2},\;(x_{i})^{2},\;\frac{t(x_{i})^{2}}{r},\;\frac{t^{5}}{r^{3}},\;\frac{t^{3}(x_{i})^{2}}{r^{3}},\;\frac{t(x_{i})^{4}}{r^{3}},\;\frac{t(x_{i})^{2}(x_{j})^{2}}{r^{3}} \; ,
\label{monoO2}\ee
giving us a total of $1+(D-1)+(D-1)+1+(D-1)+1/2(D-1)(D-2)=D(D+3)/2$. Naively we have far more coefficients to work with; e.g., in $\tilde{D}_{\mu\nu a}$ alone we have $D^{3}$ coefficients to use. However, as observed at the $\mathcal{O}(x)$ level, only a subset of the monomials and coefficients will appear. 

After much tedious algebra, one finds that the $n_{a}P^{abcd}f_{bcd}$ terms at the $\mathcal{O}(x^{2})$ level are
 \be
\begin{split}
&n_{a}P^{abcd}f_{bcd}=\frac{1}{\alpha}\biggr\{X+\frac{1}{2}P^{titj}_{\mathcal{O}(0)}\delta_{ij}\tilde{D}_{ttt}-\frac{1}{2}P^{tijk}_{\mathcal{O}(0)}\tilde{D}_{ttk}+\frac{1}{3}(D'_{tttt}P^{titj}_{\mathcal{O}(0)}\delta_{ij}+D'_{tttk}P^{kitj}_{\mathcal{O}(0)}\delta_{ij})\biggr\}\frac{t^{3}}{r}\\
&+\frac{1}{\alpha}\biggr\{Y^{ii}+\frac{1}{2}P^{tiik}_{\mathcal{O}(0)}\tilde{D}_{ttk}-\frac{1}{2}P^{titi}_{\mathcal{O}(0)}\tilde{D}_{ttt}-\frac{1}{3}\left(D'_{tttt}P^{titi}_{\mathcal{O}(0)}+D'_{tttk}P^{kiti}_{\mathcal{O}(0)}\right)\biggr\}\frac{(x_{i})^{2}t^{3}}{r^{3}}\\
&+\frac{1}{\alpha}\biggr\{Z^{iikk}-\frac{1}{2}\tilde{D}^{kk}_{\;\;\;t}P^{titi}_{\mathcal{O}(0)}-2\tilde{D}^{ki}_{\;\;\;t}P^{titk}_{\mathcal{O}(0)}\\
&-2\left(D'^{kk}_{\;\;\;tt}P^{titi}_{\mathcal{O}(0)}+2D'^{ik}_{\;\;\;tt}P^{titk}_{\mathcal{O}(0)}+D'^{kk}_{\;\;\;t\ell}P^{\ell iti}_{\mathcal{O}(0)}+2D'^{ik}_{\;\;\;t\ell}P^{\ell itk}_{\mathcal{O}(0)}\right)\biggr\}\frac{(x_{k})^{2}(x_{i})^{2}t}{r^{3}}\\
&+\frac{1}{\alpha}\left(\mathcal{X}-P^{tijk}_{\mathcal{O}(0)}\tilde{D}_{ijk}-P^{titj}_{\mathcal{O}(0)}(\tilde{D}_{itj}-\tilde{D}_{ijt})\right)rt+\frac{1}{\alpha}\biggr\{W^{kk}+P^{kjk\ell}_{\mathcal{O}(0)}\tilde{D}_{tj\ell}\\
&+P^{kji\ell}_{\mathcal{O}(0)}\delta_{ij}\tilde{D}^{k}_{\;t\ell}-P^{tktk}_{\mathcal{O}(0)}\tilde{D}_{ttt}-(P^{tkij}_{\mathcal{O}(0)}+P^{tikj}_{\mathcal{O}(0)})\tilde{D}^{k}_{\;ij}-P^{tktj}_{\mathcal{O}(0)}(\tilde{D}^{k}_{\;tj}-\tilde{D}^{k}_{\;jt})\\
&+\frac{1}{2}P^{titj}_{\mathcal{O}(0)}\delta_{ij}\tilde{D}^{kk}_{\;\;\;t}+2\left(D'^{kk}_{\;\;\;tt}P^{titj}_{\mathcal{O}(0)}\delta_{ij}+D'^{kk}_{\;\;\;t\ell}P^{\ell itj}_{\mathcal{O}(0)}\delta_{ij}\right)\biggr\}\frac{(x_{k})^{2}t}{r}\;,
\end{split}
\ee
where $X, Y^{ii}, Z^{iikk}, \mathcal{X}$, and $W^{kk}$ are some messy collection of constants independent of the $\tilde{D}$ and $D'$ coefficients. 

From counting one finds that there are more than enough coefficients to remove all of the undesired monomial expressions for arbitrary theories of gravity, and, even in the most restrictive case of $(2+1)$-dimensional $f(R)$ gravity, we will still find that we have just enough coefficients to remove all of the undesired monomials. 

To see how even the most restrictive case is satisfied, it suffices to study only a single contribution from $n_{a}^{\mathcal{O}(1)}P^{abcd}_{\mathcal{O}(0)}f^{\mathcal{O}(1)}_{bcd}$,
\be
\begin{split}
n^{\mathcal{O}(1)}_{a}P^{abcd}_{\mathcal{O}(0)}f^{\mathcal{O}(1)}_{bcd}&=-\frac{t}{\alpha}\left[P^{tijk}_{\mathcal{O}(0)}f^{\mathcal{O}(1)}_{ijk}+P^{titj}_{\mathcal{O}(0)}(f^{\mathcal{O}(1)}_{itj}-f^{\mathcal{O}(1)}_{ijt})\right]\\
&+\frac{x_{i}}{\alpha}\left[P^{ijk\ell}_{\mathcal{O}(0)}f^{\mathcal{O}(1)}_{jk\ell}+P^{itkt}_{\mathcal{O}(0)}(f^{\mathcal{O}(1)}_{tkt}-f^{\mathcal{O}(1)}_{ttk})+P^{ijtk}_{\mathcal{O}(0)}(f^{\mathcal{O}(1)}_{jtk}-f^{\mathcal{O}(1)}_{jkt})\right]\;.
\end{split}
\ee
In particular, we need only study the first line. After much algebra we find
  \be
 \begin{split}
& -\frac{t}{\alpha}P^{titj}_{\mathcal{O}(0)}(f^{\mathcal{O}(1)}_{itj}-f_{ijt}^{\mathcal{O}(1)})=\frac{1}{\alpha}\left[\mathcal{F}-P^{titj}_{\mathcal{O}(0)}(\tilde{D}_{itj}-\tilde{D}_{ijt})\right]rt\\
&+\frac{1}{2\alpha}\tilde{D}_{ttt}P^{titj}_{\mathcal{O}(0)}\delta_{ij}\frac{t^{3}}{r}-\frac{1}{2\alpha}P^{titi}_{\mathcal{O}(0)}\tilde{D}_{ttt}\frac{(x_{i})^{2}t^{3}}{r^{3}}-\frac{1}{\alpha}\left[\mathcal{M}^{kk}+P^{tktj}_{\mathcal{O}(0)}(\tilde{D}^{k}_{\;tj}-\tilde{D}^{k}_{\;jt})-\frac{1}{2}P^{titj}_{\mathcal{O}(0)}\delta_{ij}\tilde{D}^{kk}_{\;\;\;t}\right]\frac{(x_{k})^{2}t}{r}\\
&-\frac{1}{2\alpha}(\tilde{D}^{kk}_{\;\;t}P^{titi}_{\mathcal{O}(0)}+4\tilde{D}^{ki}_{\;\;\;t}P^{titk}_{\mathcal{O}(0)})\frac{(x_{k})^{2}(x_{i})^{2}t}{r^{3}}\;,
 \end{split}
 \label{n1p0f1-2exp}\ee
 where we have defined
 \be
 \mathcal{M}^{kk}\equiv \frac{4}{3}P^{titj}_{\mathcal{O}(0)}R_{i\;\;\;j}^{\;kk}(p)\,,\quad \mathcal{F}\equiv P^{titj}_{\mathcal{O}(0)}(R_{titj}(p)-R_{tijt}(p))\;.
 \ee
 Consider a $(2+1)$-dimensional spacetime. We immediately see that
 \be
 \frac{1}{2\alpha}\tilde{D}_{ttt}P^{titj}_{\mathcal{O}(0)}\delta_{ij}\frac{t^{3}}{r}-\frac{1}{2\alpha}P^{titi}_{\mathcal{O}(0)}\tilde{D}_{ttt}\frac{(x_{i})^{2}t^{3}}{r^{3}}
 \ee
 cancel each other. This is fine as it only depends on a single coefficient $\tilde{D}_{ttt}$. We have 
 \be
 \begin{split}
 \frac{1}{\alpha}\left[\mathcal{F}-P^{titj}_{\mathcal{O}(0)}(\tilde{D}_{itj}-\tilde{D}_{ijt})\right]rt&=\frac{1}{\alpha}\left[\mathcal{F}-P^{titi}_{\mathcal{O}(0)}\left(\tilde{D}_{xtx}-\tilde{D}_{xxt}+\tilde{D}_{yty}-\tilde{D}_{yyt}\right)\right]rt\;,
 \end{split}
 \ee
 \be
 \begin{split}
 -\frac{1}{2\alpha}(\tilde{D}^{kk}_{\;\;t}P^{titi}_{\mathcal{O}(0)}+4\tilde{D}^{ki}_{\;\;\;t}P^{titk}_{\mathcal{O}(0)})\frac{(x_{k})^{2}(x_{i})^{2}t}{r^{3}}&=-\frac{1}{2\alpha}\biggr\{5\tilde{D}_{xxt}x^{4}+5\tilde{D}_{yyt}y^{4}+(\tilde{D}_{xxt}+\tilde{D}_{yyt})x^{2}y^{2}\biggr\}\frac{t}{r^{3}}\;,
 \end{split}
 \ee
 and
 \be
 \begin{split}
& -\frac{1}{\alpha}\left[\mathcal{M}^{kk}+P^{tktj}_{\mathcal{O}(0)}(\tilde{D}^{k}_{\;tj}-\tilde{D}^{k}_{\;jt})-\frac{1}{2}P^{titj}_{\mathcal{O}(0)}\delta_{ij}\tilde{D}^{kk}_{\;\;\;t}\right]\frac{(x_{k})^{2}t}{r}=-\frac{1}{\alpha}\left(\frac{4}{3}P^{titi}_{\mathcal{O}(0)}R_{yxxy}(p)\right)rt\\
 &-\frac{1}{\alpha}P^{titi}_{\mathcal{O}(0)}\left[(\tilde{D}_{xtx}-\tilde{D}_{xxt})\frac{x^{2}t}{r}+(\tilde{D}_{yty}-\tilde{D}_{yyt})\frac{y^{2}t}{r}\right]\;.
 \end{split}
 \ee
 Let us now set $\tilde{D}_{kkt}=0$. This choice yields the two expressions
 \be
 \begin{split}
 \frac{1}{\alpha}\left[\mathcal{F}-P^{titj}_{\mathcal{O}(0)}(\tilde{D}_{itj}-\tilde{D}_{ijt})\right]rt&=\frac{1}{\alpha}\left[\mathcal{F}-P^{titi}_{\mathcal{O}(0)}\left(\tilde{D}_{xtx}+\tilde{D}_{yty}\right)\right]rt
 \end{split}
 \ee
 and
 \be
 \begin{split}
& -\frac{1}{\alpha}\left[\mathcal{M}^{kk}+P^{tktj}_{\mathcal{O}(0)}(\tilde{D}^{k}_{\;tj}-\tilde{D}^{k}_{\;jt})-\frac{1}{2}P^{titj}_{\mathcal{O}(0)}\delta_{ij}\tilde{D}^{kk}_{\;\;\;t}\right]\frac{(x_{k})^{2}t}{r}\\
&=-\frac{1}{\alpha}\left(\frac{4}{3}P^{titi}_{\mathcal{O}(0)}R_{yxxy}(p)\right)rt-\frac{1}{\alpha}P^{titi}_{\mathcal{O}(0)}\left[\tilde{D}_{xtx}\frac{x^{2}t}{r}+\tilde{D}_{yty}\frac{y^{2}t}{r}\right]\;.
 \end{split}
 \ee
 Let us further choose that $\tilde{D}_{xtx}=\tilde{D}_{yty}\equiv\tilde{D}$. The second expression then becomes
 \be
 -\frac{1}{\alpha}\left(\frac{4}{3}P^{titi}_{\mathcal{O}(0)}R_{yxxy}(p)\right)rt-\frac{1}{\alpha}P^{titi}_{\mathcal{O}(0)}\tilde{D}rt\;.
 \ee
 Defining $4/3P^{titi}_{\mathcal{O}(0)}R_{yxxy}(p)\equiv\mathcal{M}$, we find that the following combination must be made to vanish:
 \be
 -\frac{1}{\alpha}\left[\mathcal{M}-\mathcal{F}+3P^{titi}_{\mathcal{O}(0)}\tilde{D}\right]rt
 \ee
We have the freedom to choose $\tilde{D}$ such that this monomial vanishes. 
 
 The reason this specific case is enough to show that there are enough coefficients to remove all of the $\mathcal{O}(x^{2})$ monomial contributions to $\int_{\Sigma}n_{a}P^{abcd}f_{bcd}$ is that every type of possible monomial is present. Any additional contributions which come into play can easily be handled by (i) altering the choice of $\tilde{D}_{\mu\nu a}$, and (ii) having the presence of $\tilde{D}'_{\mu\nu\rho a}$ coefficients. The only monomial which might give us pause is that proportional to $t(x_{i})^{2}/r$, as the $\tilde{D}_{ttt}$ happened to exactly cancel. It turns out, however, that there are enough $D'$ coefficients to deal with these monomials.
 
 In summary, by modifying $\xi_{a}$ and $n_{a}$, we have more than enough coefficients to remove all of the monomial contributions to $n_{a}P^{abcd}f_{bcd}$ that do not vanish due to integration over the sphere, through the $\mathcal{O}(x^{2})$ level. Therefore, while there might be $\mathcal{O}(x^{3})$ contributions to the integrand, these terms are sufficiently smaller than those we wish to keep in the equations of motion, allowing us to effectively neglect the undesired contribution $\int_{\Sigma}n_{a}P^{abcd}f_{bcd}$. 
 
 \subsection*{Eliminating $q^{a}$}
\indent
Last, let us discuss how to eliminate another unwanted term,
\be
-\frac{1}{4G\hbar}\int_{\Sigma}dAd\tau n_{a}q^{a} \; ,
\ee
where $q^{a}=\nabla_{b}(P^{adbc}+P^{acbd})\nabla_{c}\xi_{d}$. This term is only present for non-Lovelock theories of gravity, such as non-Einstein $f(R)$ gravity. Only the symmetric parts of $\nabla_{c}\xi_{d}$ survive the contraction. From (\ref{Killingfailure}), we see that the symmetric parts have both ${\mathcal O}(x^2)$ and ${\mathcal O}(1)$ parts. Since $n_a$ is of order $x$, the ${\mathcal O}(x^2)$ part of $q^a$ gives a term in $n_a q^a$ of order $x^3$, and we can therefore neglect it. But the ${\mathcal O}(1)$ $i-j$ contributions cannot be neglected outright:
\be
-\frac{1}{4G\hbar}\int_{\Sigma}d\Sigma_{a}\nabla_{b}(P^{aibj})(\nabla_{i}\xi_{j}+\nabla_{j}\xi_{i})\;.
\label{failkillingeqn}
\ee
To match our approximations we must therefore eliminate this contribution for non-Lovelock theories of gravity. This is indeed possible, as we now show. Because of the form, Eq. (\ref{Killingfailure}), of $\nabla_{(i}\xi_{j)}$, terms with $i\neq j$ integrate to zero in
(\ref{failkillingeqn}). When $i=j$, the integrand is of $\mathcal{O}(x)$ for the combination $n^{(1)}_{t}(\nabla_b P^{tibi}_{\mathcal{O}(0)})\nabla_{i}\xi_{i}$. This yields two types of monomials:
\be
 \frac{t^{2}}{r},\quad \frac{t^{2}(x_{i})^{2}}{r^{3}}\;.
\ee
However, precisely these monomials already appear in (\ref{monoO1}). They can therefore be absorbed in the $\mathcal{O}(x)$ contributions to $n_{a}P^{abcd}f_{bcd}$ that have already been shown to be eliminated; the counting argument discussed at length above is not altered. 
The integrand of (\ref{failkillingeqn}) will be of ${\mathcal O}(x^2)$ in two ways: (i) $n_{a}^{(2)}(\nabla_{b}P^{aibj})^{(0)}\nabla_{(i}\xi_{j)}$, or (ii) $n_{a}^{(1)}(\nabla_{b}P^{aibj})^{(1)}\nabla_{(i}\xi_{j)}$. Together, the only monomials that appear are
\be
\frac{t^{3}}{r},\quad \frac{t^{3}(x_{i})^{2}}{r^{3}}, \quad \frac{t(x_{i})^{2}}{r},\quad \frac{t(x_{i})^{2}(x_{j})^{2}}{r^{3}}
\ee
matching the monomials already appearing in (\ref{monoO2}). In summary, the terms appearing in (\ref{failkillingeqn}) can be readily eliminated by the coefficients we use to dispose of similar terms in $n_{a}P^{abcd}f_{bcd}$, without altering the counting.

\newpage

\section*{APPENDIX B: EQUATING INTEGRANDS}
\noindent
We have seen that Clausius' theorem, $Q = \Delta S_{\rm rev}/T$, leads to an equality between integrals of the form
\be
\int_{\Sigma}dAd \tau A_{ab}\xi^{a}n^{b}=\int_{\Sigma}dAd \tau T_{ab}\xi^{a}n^{b} \; . \label{equalintegrals}
\ee
For Einstein gravity, $A_{ab} = \frac{1}{8 \pi G} R_{ab}$, while for general theories of gravity, $A_{ab}$ can be read off from the left-hand side of (\ref{Clausiusgeneral}).
In this appendix, we show that the equality of integrals (\ref{equalintegrals}) implies the equality of their integrands:
\be
A_{ab} \xi^{a}n^{b}=T_{ab}\xi^{a}n^{b}\;.
\ee
Ordinarily, the equality of integrands follows from the equality of integrals if the boundaries of the domain of integration can be suitably varied without affecting the equality of the integrals. 

Defining the symmetric matrix $M_{ab}\equiv A_{ab}-T_{ab}$, and with the proper time element on the hyperboloid given by $d \tau = dt \alpha/r$, we can write (\ref{equalintegrals}) as
\be
0=\int^{\epsilon}_{0}dt \frac{\alpha}{r(t)} \int_{\omega(t)}dAM_{ab}\xi^{a}n^{b}\;.
\ee
We would like to conclude from this that $M_{ab} \xi^a n^b = 0$. Because $\epsilon$ is arbitrary, for this integral to vanish for all values of $\epsilon$, the standard argument from calculus implies that the integrand must itself be zero:
\be
0= \int_{\omega(t)} dA M_{ab}\xi^{a}n^{b} \; ,
\ee
for all spheres $\omega(t)$. However, we cannot apply the same argument to this integral because a sphere has no boundary to vary.

Expanding the integrand gives
\be
 0=\int dA\left[M_{00}rt+M_{0i}tx^{i}\left(1+\frac{t}{r}\right)+M_{ii}\frac{t(x^{i})^{2}}{r}+M_{ij,i\neq j}\frac{tx^{i}x^{j}}{r}\right]\;.
 \ee
Integration over the sphere causes the terms in the integrand proportional to odd powers of $x^i$ to automatically vanish, telling us nothing about $M_{ij,i\neq j}$ and $M_{0i}$. We see, however, that the other components must obey the condition
\be
M_{00}+\frac{1}{(D-1)}\sum_{i}M_{ii}=0\;.\label{cond1}
\ee

To proceed, note that (\ref{equalintegrals}) also holds for a different hyperboloid, $\Sigma'$, obtained by an active Lorentz transformation of $\Sigma$. This active transformation does not affect the matrix $M$, whose elements are evaluated at $p$, but transforms the vectors $\xi$ and $n$ to $\xi'$ and $n'$. We then follow this with a passive Lorentz transformation on the coordinates such that the components of the new $\xi'$ and $n'$ are the same as the original components of the old $\xi$ and $n$. Under a passive Lorentz transformation, $M$ transforms as a matrix, and we have
\be
 0=\int_{\Sigma'}dAdt\frac{\alpha}{r} M'_{ab}\xi^{a}n^{b} \Rightarrow 0 = \int dA\left[M'_{00}rt+M'_{0i}tx^{i}\left(1+\frac{t}{r}\right)+M'_{ii}\frac{t(x^{i})^{2}}{r}+M'_{ij,i\neq j}\frac{tx^{i}x^{j}}{r}\right]
 \ee
from which we find
\be
M'_{00}+\frac{1}{(D-1)}\sum_{i}M'_{ii}=0\;. \label{cond2}
\ee
We now show that (\ref{cond1}) and (\ref{cond2}) are enough to claim $M_{ab}\propto\eta_{ab}$. Perform a Lorentz transformation in the $0-1$ plane. Then applying (\ref{cond1}) and (\ref{cond2}) leads to
\be
M_{00}=-M_{11}-\frac{2\beta\gamma^{2}}{(1-\gamma^{2})}M_{01}
\ee
For this to hold for all $\beta$, we conclude that $M_{01}=0$. Moreover, $M_{00}=-M_{11}$. A similar argument holds for Lorentz boosts in other planes, and therefore, $M_{00}=-M_{11}=-M_{22}=...$, and $M_{0i}=0$. It is also straightforward to show that $M_{ij}=0$ for $i\neq j$ by first performing a rotation on $M_{ab}$, and then a Lorentz boost. In summary, we find that $M_{ab}$ is a diagonal matrix with $M_{00}=-M_{ii}$. Hence $M_{ab}\propto\eta_{ab}$.
But since $\eta_{ab} \xi^a n^b = 0$, we find 
\be
 M_{ab}\xi^{a}n^{b}=0
 \ee
as desired.

\end{document}